\documentclass[a4paper,12pt]{article}
\pdfoutput=1

\usepackage{jheppub} 

\usepackage{amsmath}
\usepackage{amssymb}
\usepackage{lmodern}
\usepackage{accents}
\usepackage{booktabs}
\usepackage{caption, subcaption}
\usepackage{mathrsfs}

\usepackage{float}

\usepackage[T1]{fontenc} 

\usepackage{tikz}
\usetikzlibrary{calc,arrows,decorations.markings,matrix,decorations.pathreplacing,positioning,fit, arrows}
\newcommand{\mr}[1]{\mathrm{#1}}
\newcommand{\mf}[1]{\mathfrak{#1}}
\newcommand{\ms}[1]{\mathsf{#1}}
\newcommand{\mc}[1]{\mathcal{#1}}
\newcommand{\mb}[1]{\mathbb{#1}}

\newcommand{\f}[2]{\frac{#1}{#2}}

\newcommand{\no}[1]{:\mathrel{\mspace{2mu}#1\mspace{2mu}}:}
\newcommand{\nn}{\nonumber}

\newcommand{\jt}[1]{\theta_1\!\l(\tau \middle| #1 \r)}
\newcommand{\af}[1]{\hat{a}\l( #1 \r)}
\newcommand{\pe}[3][]{\ensuremath{\mathtt{PE}^{#1}_{#2}\l[ #3 \r]}}

\let\l\relax
\let\r\relax
\let\d\relax
\newcommand{\l}{\left}
\newcommand{\r}{\right}
\newcommand{\d}{\mr{d}}
\newcommand{\dts}{\!\ifmmode\mathinner{\ldotp\kern-0.2em\ldotp\kern-0.2em\ldotp}\else.\kern-0.13em.\kern-0.13em.\fi\!}
\let\i\relax
\newcommand{\i}{\indices}

\DeclareMathOperator{\diag}{diag}
\DeclareMathOperator*{\res}{Res}

\DeclareMathOperator{\tr}{Tr}
\DeclareMathOperator{\Tr}{Tr}
\DeclareMathOperator{\im}{\mathbb{I}m}

\newcommand{\cF}{\mathcal{F}}

\newcommand{\cN}{\mathcal{N}}

\newcommand{\bC}{\mathbb{C}}

\newcommand{\bZ}{\mathbb{Z}}

\newcommand\SU{\mathrm{SU}}
\newcommand\U{\mathrm{U}}

\newcommand{\matht}[1]{\ensuremath{\boldsymbol{#1}}}
\newcommand{\be}{\begin{equation}}
\newcommand{\ee}{\end{equation}}
\newcommand{\bea}{\begin{equation}\begin{aligned}}
\newcommand{\eea}{\end{aligned}\end{equation}}
\newcommand{\ie}{\textit{i.e.}}
\newcommand{\eg}{\textit{e.g.}}
\newcommand{\rep}[1]{\ensuremath{\boldsymbol{#1}}}
\newcommand{\wt}{\widetilde}
\newcommand{\wb}{\overline}

\newcommand{\bullup}{\accentset{\bullet}}

\title{Elliptic non-Abelian Donaldson-Thomas invariants of \matht{\bC^3}}

\author{Francesco Benini,}
\author{Giulio Bonelli,}
\author{Matteo Poggi}
\author{and Alessandro Tanzini}
\affiliation{International School of Advanced Studies (SISSA/ISAS) and INFN Sezione di Trieste \\
   via Bonomea 265, I-34136 Trieste, Italy}

\emailAdd{fbenini}
\emailAdd{bonelli}
\emailAdd{mpoggi}
\emailAdd{tanzini@sissa.it}
\preprint{SISSA~29/2018/FISI}

\abstract{%
We compute the elliptic genus of the D1/D7 brane system in flat space, finding a non-trivial dependence on the number of D7 branes, and provide an \mbox{F-theory} interpretation of the result. We show that the JK-residues contributing to the elliptic genus are in one-to-one correspondence with coloured plane partitions and that 
the elliptic genus can be written as a chiral correlator of vertex operators on the torus. We also study the quantum mechanical system describing D0/D6 bound states on a circle, which leads to a plethystic exponential formula that can be connected to the M-theory graviton index on a multi-Taub-NUT background. The formula is a conjectural expression for higher-rank equivariant K-theoretic Donaldson-Thomas invariants on $\mathbb{C}^3$.
}

\begin{document}
 \maketitle

\section{Introduction}

The study of brane dynamics has revealed, over the years, to be a constant source of delightful results both in physics and mathematics. It offers valuable insights into the non-perturbative dynamics of gauge and string theories, and it displays deep connections with enumerative geometry via BPS bound-state counting. Often brane systems provide a string theory realisation of interesting moduli spaces, and supersymmetric localisation allows us to perform the exact counting of BPS states in a variety of them.

This philosophy has been applied successfully in many contexts. For instance, the $S^2$ partition functions \cite{Benini:2012ui, Doroud:2012xw} of gauged linear sigma models (GLSMs) capture geometric properties of the moduli spaces of genus-zero pseudo-holomorphic maps to the target, and represent a convenient way to extract Gromov-Witten invariants \cite{Jockers:2012dk}. They show that suitable coordinates enjoy mutations of cluster algebras \cite{Benini:2014mia}, as physically suggested by IR dualities \cite{Benini:2011mf}. As another example, certain equivariant K-theories of vortex moduli spaces are conveniently captured by a twisted 3D index \cite{Gukov:2015sna, Benini:2015noa}. Such an object is intimately related to black hole entropy in AdS$_4$ \cite{Benini:2015eyy, Benini:2016rke}, thus providing a sort of generalisation of Gopakumar-Vafa invariants \cite{Gopakumar:1998jq}.

Exact $S^2$ partition functions have been exploited in the study of D1/D5 brane systems in
\cite{Bonelli:2013rja,Bonelli:2013mma} providing a direct link between quantum cohomologies of \mbox{Nakajima} quiver varieties,
quantum integrable systems of hydrodynamical type, and higher-rank equivariant Donaldson-Thomas
invariants of $\mathbb{P}^1\times \mathbb{C}^2$ \cite{Bonelli:2014iza, Bonelli:2015kpa}. A BPS state counting for the D0/D2 brane system analogous to the one considered in this paper was performed in \cite{Poggi:2017kut}, providing an elliptic generalisation of vortex counting results \cite{Bonelli:2011fq, Bonelli:2011wx}.

In this paper, we analyse the D1/D7 brane system on an elliptic curve in type IIB superstring theory. The effective dynamics of the D1-branes is captured by a two-dimensional $\mathcal{N}=(2,2)$ supersymmetric GLSM living on the elliptic curve, and whose classical vacua describe the moduli space of rank-$N$ sheaves on $\mathbb{C}^3$, where $N$ is the number of D7-branes. The supersymmetric partition function of this theory computes the elliptic genus of the above moduli space. We also analyse the dimensionally reduced cases of D0/D6 and D($-1$)/D5 branes, which compute the generalised Witten index and the equivariant volume of the same moduli space, respectively.

The last two cases were extensively studied for rank one, in view of their relation with black-hole entropy, microstate counting \cite{Ooguri:2004zv} and Donaldson-Thomas (DT) invariants \cite{Donaldson:1996kp}. The latter are in turn mapped to Gromov-Witten invariants by the MNOP relation \cite{MNOP06a, MNOP06b}. Less is known in the higher-rank case,%
\footnote{The higher-rank D0/D6 partition function for \emph{compact} Calabi-Yau three-folds, related to DT invariants of \emph{unframed} sheaves, was computed in \cite{Toda:2009, Stoppa:2012, Manschot:2010qz}. It does not factorize as the $N$-th power of the Abelian case.}
except for the D($-1$)/D5 system whose partition function was conjectured to factorise as the $N$-th power of the Abelian one \cite{Awata:2009dd, Nekrasov:2009jap}. In this paper we provide evidence for such a factorisation conjecture. 

On the other hand, we find that the elliptic genus and the generalised Witten index do not factorize and give new interesting results. In Proposition 5.1 of \cite{Nekrasov:2014nea}, a relation between the higher-rank equivariant K-theoretic DT invariants on a three-fold $X$ and the M2-brane contribution to the M-theory index on a $A_{N-1}$ surface fibration over $X$ was established. A conjectural plethystic exponential form for the equivariant K-theoretic DT invariants in higher rank was proposed in \cite{Awata:2009dd} for the case $X = \bC^3$. In this paper we confirm that proposal. For rank one, the D0/D6 system on a circle is known to compute the eleven-dimensional supergravity index, which can indeed be expressed in an elegant plethystic exponential form \cite{Nekrasov:2009jap}. We show that the same is true in the higher-rank case. In fact, extending the construction of \cite{Townsend:1995kk}, the \mbox{M-theory} lift of the D0/D6 system in the presence of an Omega background is given by a $\mathrm{TN}_N \times \mathbb{C}^3$ fibration over a circle \cite{Nekrasov:2009jap}, where $\mathrm{TN}_N$ is a multi-center Taub-NUT space and whose charge $N$ equals the number of D6-branes. The fibration is such that the fiber space is rotated by a $U(1)^3$ action as we go around the circle. The multi-center Taub-NUT space looks asymptotically as a lens space $S^3/\mathbb{Z}_N \times \mathbb{R}^+$, precisely as the asymptotic behaviour of the $A_{N-1}$ surface singularity $\mathbb{C}^2/\mathbb{Z}_N$. This implies the appearance in the higher-rank index of twisted sectors carrying irreducible representations of the cyclic group, which spoils the factorisation property.   

In the elliptic case---describing the D1/D7 system---a novelty appears: because of anomalies in the path integral measure, there are non-trivial constraints on the fugacities of the corresponding symmetries. Once these constraints are taken into account, the higher-rank elliptic index takes a particularly simple form, which can be traced back to a suitable geometric lift to F-theory \cite{Vafa:1996xn}.

We use supersymmetric equivariant localisation to evaluate the elliptic genus: this reduces the computation to a residue problem with Jeffrey-Kirwan contour prescription \cite{Benini:2013nda, Benini:2013xpa}. As we discuss in the following, some subtleties arise due to degenerate and higher-order poles. We implement a desingularisation procedure, whose final result is a classification of the poles in terms of (coloured) plane partitions.

Finally, we propose a realisation of the elliptic genus as a chiral correlator of free fields on the torus---with the aim of exploring the underlying integrable structure in the spirit of the BPS/CFT correspondence \cite{Nekrasov:2015wsu}. 

The content of the paper is as follows. In Section~\ref{sec: Abelian case} we compute the elliptic genus of the D1/D7 system in the rank-one case, as well as its dimensional reductions to the trigonometric and rational cases. We review the plethystic formula describing the latter. In Section~\ref{sec: non-Abelian case} we address the higher-rank case. We first provide evidence for the factorisation conjecture in the rational case, and then we study a conjectural plethystic exponential form for the trigonometric case in equation \eqref{eq:facttrig}. The elliptic genus is displayed in equation \eqref{eq:ZkN}. Subsections \ref{sec: M-theory} and \ref{sec: F-theory} contain respectively comments on the M-theory and F-theory interpretations of our results. Section~\ref{sec: free field} describes the free-field realisation of the elliptic genus. Section~\ref{sec: conclusions} is devoted to conclusions and open questions. Many technical details are relegated to the appendices.

\section{Elliptic DT invariants of $\matht{\bC^3}$: Abelian case}
\label{sec: Abelian case}

To study (equivariant) Donaldson-Thomas invariants \cite{Donaldson:1996kp} of a three-fold, one can employ a string theory brane construction \cite{MNOP06a,MNOP06b}. In particular, in order to study the Hilbert scheme of points on the three-fold we place a single Euclidean D5-brane on the three-fold, and some number $k$ of D($-1$)-branes on its worldvolume. In order to preserve supersymmetry (SUSY), a certain $B$-field must be turned on along the D5-brane \cite{Witten:2000mf}. This creates a trapping potential that confines the D($-1$)-branes on the D5-brane worldvolume. At this point, the supersymmetric theory on the D($-1$)-branes---which is a matrix model---contains information about the sought-after invariants. Much information can be extracted with supersymmetric field theory techniques.

We are interested in the simplest case that the three-fold is $\bC^3$ (the same ideas apply to three-folds with richer topology). In fact, we can similarly study K-theoretic and elliptic generalisations of the DT invariants by adding one or two directions to the brane setup. Specifically, we can study a D6-brane wrapped on the three-fold and $k$ D0-branes on its worldvolume: the quantum mechanics on the D0-branes captures the K-theoretic DT invariants of the three-fold \cite{Okounkov:2015spn}. Besides, we can study a D7-brane wrapped on the three-fold and $k$ D1-branes on its worldvolume: the two-dimensional theory on the D1-branes allows us to define ``elliptic DT invariants'' of the three-fold. We define them as the elliptic genera of the Hilbert schemes of $k$ points on the three-fold. From the QFT point of view, they are the elliptic genera of the theories living on the D1-branes.

While in this section we study the D1/D7 system with a single D7-brane, in Section~\ref{sec: non-Abelian case} we will move to higher-rank DT invariants. They are captured by the D1/D7 system with $N$ multiple D7-branes wrapping the three-fold (here $\bC^3$). This will define for us ``elliptic non-Abelian DT invariants''.

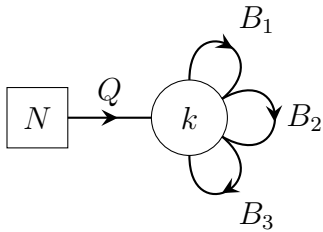
\begin{figure}
\centering
\begin{tikzpicture}
\draw (-.4,-.4) rectangle (.4, .4) node[midway] {$N$};
\draw (2,0) circle [radius = .5] node {$k$};
\draw [thick, decoration = {markings, mark=at position .6 with {\arrow[scale=1.5]{stealth}}}, postaction=decorate] (.4,0) to node[midway, above] {$Q$} (1.5,0) ;
\draw [thick, decoration = {markings, mark=at position .55 with {\arrow[scale=1.5, rotate=14]{stealth}}}, postaction=decorate] (2,0.5) to [out = 90, in = 40, looseness=5] node [midway, above right] {$B_1$} (2.43, 0.25);
\draw [thick, decoration = {markings, mark=at position .45 with {\arrowreversed[scale=1.5, rotate=14]{stealth}}}, postaction=decorate] (2,-0.5) to [out = -90, in = -40, looseness=5] node [midway, below right] {$B_3$} (2.43, -0.25);
\draw [thick, decoration = {markings, mark=at position .53 with {\arrow[scale=1.5, rotate=10]{stealth}}}, postaction=decorate] (2.43,0.25) to [out = 30, in = -30, looseness=5.5] node [midway, right] {$B_2$} (2.43, -0.25);
\end{tikzpicture}
\caption{2d $\mc{N}=(2,2)$ quiver gauge theory with a $\mr{U}(k)$ vector multiplet; $Q$, $B_{a=1,2,3}$ chiral multiplets; $\mr{SU}(N)$ flavour symmetry (in Section~\ref{sec: Abelian case} we take $N=1$). The superpotential is $W=\tr\l(B_1[B_2,B_3]\r)$.
\label{fig: quiver}}
\end{figure}

The 2d theory living on $k$ D1-branes probing $N$ D7-branes has $\cN=(2,2)$ supersymmetry and is described by the quiver diagram in Figure~\ref{fig: quiver}. The field content is given by a $\U(k)$ vector multiplet, three chiral multiplets $B_{a=1,2,3}$ in the adjoint representation and $N$ chiral multiplets $Q_\alpha$ in the fundamental representation. Moreover there is a superpotential
\be
W = \Tr \big( B_1 [ B_2, B_3] \big) \;.
\ee
Besides the $\U(k)$ gauge symmetry, the theory has $\SU(N)$ flavour symmetry acting on the $N$ chiral multiplets $Q_\alpha$ in the antifundamental representation and $\U(1)^2$ flavour symmetry acting on $B_a$. At the classical level there is $\U(1)_L \times \U(1)_R$ R-symmetry,%
\footnote{When the theory is superconformal, the superconformal R-charges can be computed with \mbox{$c$-extremisation} \cite{Benini:2012cz, Benini:2013cda}.}
however in the quantum theory the anomaly breaks the (anti-diagonal) axial part to $\bZ_N$. This is related to the fact that the theory is not conformal, rather it is gapped with a dynamically generated scale.

We can associate fugacities to the Cartan generators of the gauge, flavour and $\U(1)_L$ symmetry groups, as summarised in Table~\ref{tab: holo}. We express the fugacities as exponentials of chemical potentials, \eg{}, $y=e^{2\pi i z}$. As we will see, it is convenient to define the variables
\be
\label{def epsilon_i}
\epsilon_1 = \tfrac13\epsilon + \zeta_1 \;,\qquad \epsilon_2 = \tfrac13 \epsilon + \zeta_2 - \zeta_1 \;,\qquad \epsilon_3 = \tfrac13 \epsilon - \zeta_2
\ee
that satisfy the relation
\be
\label{relation epsilon_i}
\epsilon_1 + \epsilon_2 + \epsilon_3 = \epsilon \;.
\ee
Because of the anomaly, we should restrict to $\epsilon \in \bZ/N$. Notice that fugacities are invariant under shift of the chemical potentials by $1$, however, because of 't~Hooft anomalies, partition functions in general are not.

\begin{table}[t]
\centering
\begin{tabular}{lccccc}
\toprule
Group & $B_1$ & $B_2$ & $B_3$ & $Q$ & Fugacity \\
\midrule
$\mr{U}(k)$ & adj & adj & adj & fund & $e^{2\pi i u_i} $ \\
$\mr{SU}(N)$ & \rep{1} & \rep{1} & \rep{1} & anti-fund & $ e^{- 2\pi i z_\alpha} $\\
$\mr{U}(1)_1 $ & 1 & $-1$ & 0 & 0 & $e^{2\pi i \zeta_1}$ \\
$\U(1)_2$ & 0 & 1 & $-1$ & 0 & $e^{2\pi i \zeta_2}$ \\
$\U(1)_L$ & $\frac13$ & $\frac13$ & $\frac13$ & 0 & $e^{2\pi i \epsilon}$ \\
\bottomrule
\end{tabular}
\caption{Gauge, flavour and R- symmetry groups, charges of chiral multiplets and associated fugacities (exponentials of chemical potentials). The multiplets $B_I$ have vector-like R-charge $\frac23$ so that the left-moving R-charge is $\frac13$. The flavour symmetry fugacities are constrained to satisfy $\prod_\alpha e^{-2\pi i z_\alpha}=1$.
\label{tab: holo}}
\end{table}

We want to compute the elliptic genus \cite{Schellekens:1986yi, Schellekens:1986yj, Witten:1986bf}---\ie{} the supersymmetric index or $T^2$ partition function---of the theory. More precisely, we compute the equivariant elliptic genus, with fugacities for the global symmetries in Table~\ref{tab: holo}. In the path integral formulation, they correspond to holonomies on $T^2$ for background gauge fields%
\footnote{In order to preserve two chiral supercharges, we do not turn on a fugacity for $\U(1)_R$.}
(more details can be found in \cite{Benini:2013nda, Benini:2013xpa, Pestun:2016zxk, Benini:2016qnm}). In this section we focus on the Abelian case $N=1$. Using the formulas in \cite{Benini:2013nda, Benini:2013xpa} (see also \cite{Gadde:2013ftv}), the elliptic genus is
\begin{multline}
\label{elliptic genus N=1}
Z_k^{(1)}(\epsilon_a,\tau) = \frac{1}{k!}
\l[ \frac{ 2\pi\eta(\tau)^3 \, \jt{\epsilon_{12}} \jt{\epsilon_{13}} \jt{\epsilon_{23}} }{ \jt{\epsilon_1} \jt{\epsilon_2} \jt{\epsilon_3} \jt{\epsilon} } \r]^k
\int_\text{JK} \prod_{i=1}^k \d u_i \, \prod_{i=1}^k \f{\jt{u_i-\epsilon}}{\jt{u_i}} \times {} \\
{} \times \prod_{\substack{i,j=1\\i\neq j}}^k
\frac{ \jt{u_{ij}} \jt{u_{ij}-\epsilon_{12}} \jt{u_{ij}-\epsilon_{13}} \jt{u_{ij}-\epsilon_{23}} }{ \jt{u_{ij}+\epsilon_{1}} \jt{u_{ij}+\epsilon_{2}} \jt{u_{ij}+\epsilon_{3}} \jt{u_{ij}-\epsilon} } \;.
\end{multline}
Here $\tau$ is the modular parameter of the torus and we can define
\be
p = e^{2\pi i \tau} \;.
\ee
Then we used the short-hand notations
\be
u_{ij} \,\equiv\, u_i - u_j \;,\qquad\qquad \epsilon_{ab} \,\equiv\, \epsilon_a + \epsilon_b
\ee
as well as (\ref{def epsilon_i}) and (\ref{relation epsilon_i}). The function $\theta_1$ is a Jacobi theta function (see Appendix~\ref{app: special functions}), and we used that it is odd in the second argument. As explained in \cite{Benini:2013nda, Benini:2013xpa}, the integral is along a specific contour that corresponds to the Jeffrey-Kirwan (JK) residue \cite{JK95}.

Two comments are in order. First, the integrand in (\ref{elliptic genus N=1}) is a doubly-periodic function of $u_i$, invariant under $u_i \to u_i + a + b\tau$ for $a,b\in\mb{Z}$, only if $\epsilon \in \bZ$. For generic values of $\epsilon$, instead, the integrand picks up a phase $e^{2\pi i b \epsilon}$. This is how the gauge-R-symmetry anomaly manifests itself in the localised path-integral formulation. Thus, the elliptic genus makes sense only for those quantised values of $\epsilon$. There is also an 't~Hooft anomaly for the R-symmetry, and as a result we find
\be
\label{'t Hooft anomaly}
Z_k^{(1)}(\epsilon_1, \epsilon_2, \epsilon_3 + 1, \tau) = (-1)^k \, Z_k^{(1)}(\epsilon_1, \epsilon_2, \epsilon_3, \tau) \;.
\ee
This corresponds to the shift $\epsilon \to \epsilon+1$, $\zeta_1 \to \zeta_1 - \frac13$, $\zeta_2 \to \zeta_2 - \frac23$. Exactly the same sign is picked up if we shift one of the other $\epsilon_a$'s.

Second, the prefactor outside the integral in (\ref{elliptic genus N=1}) is ill-defined for $\epsilon\in\bZ$ because $\jt{\epsilon}=0$. To solve this conflict, we proceed as in \cite{Benini:2013nda, Benini:2013xpa}. We introduce an extra chiral multiplet $P$ in the $\det^{-1}$ representation of $\U(k)$. In the new theory, the continuous R-symmetry is non-anomalous and we can take generic values of $\epsilon$. In particular, the limit $\epsilon\to0$ is well-defined and finite. Of course, the theory with $P$ is different from the one we are interested in. However, at $\epsilon=0$ we can introduce a real mass for $P$ and remove it from the low-energy spectrum.%
\footnote{A real mass has R-charge 2, therefore it is compatible with the elliptic genus computation only at $\epsilon=0$.}
Therefore the elliptic genus of the theory without $P$ at $\epsilon =0$ is equal to the $\epsilon\to0$ limit of the elliptic genus of the theory with $P$. Notice that the one-loop determinant of $P$ satisfies $\lim_{\epsilon\to0} Z_P(u_i) = 1$. With a suitable choice of the regularisation parameter $\eta$ in the JK residue, \ie{} with a suitable choice of contour, the poles of $Z_P$ at $\epsilon\neq 0$ do not contribute to the integral. Thus---with this particular choice---the multiplet $P$ can be completely ignored: one computes the integral (\ref{elliptic genus N=1}) for generic $\epsilon$ and then takes the $\epsilon\to0$ limit. More details and examples can be found in \cite{Benini:2013nda, Benini:2013xpa}.

\subsection{Evaluation}
\label{sec: evaluation ell genus N=1}

In order to evaluate the Jeffrey-Kirwan residue integral in (\ref{elliptic genus N=1}) we follow similar examples in \cite{Benini:2013xpa}. We first identify the hyperplanes where the integrand has pole singularities:
\be
\label{eq:hyperplanes}
H_{F;i} = \{u_i = 0\} \;,\quad H_{V;ij} = \{u_i - u_j = \epsilon\} \;,\quad H_{A;ij}^a = \{u_i - u_j = - \epsilon_a\} \quad a=1,2,3 \;.
\ee
The singular hyperplanes $H_F$ are due to the one-loop determinant of the chiral multiplet $Q$, the hyperplanes $H_A$ are due to $B_a$ while the hyperplanes $H_V$ are due to vector multiplets associated to the roots of $\U(k)$. The associated charge vectors, which are the charge vectors of the chiral or vector multiplets responsible for the singularities, are:
\be\label{eq:charge_vector}
\vec{h}_{F;i} = (0, \dots, \underbrace{1}_i, \dots,0) \;, \qquad
\vec{h}_{V;ij}  = \vec{h}_{A;ij} = (0,\dots,\underbrace{1}_i,\dots,\underbrace{-1}_j,\dots,0) \;.
\ee
The poles that can contribute to the elliptic genus have maximal codimension, \ie{} they are points in the $u$-torus where $k$ linearly-independent hyperplanes meet (as we will discuss momentarily, the total number of hyperplanes through the point is in general larger than $k$). Those points are solutions to systems of linear equations
\be
\label{eq:ls}
\ms{Q}^{\ms{T}} \begin{pmatrix} u_1 \\ \vdots \\ u_k \end{pmatrix} = \begin{pmatrix} d_1 \\ \vdots \\ d_k \end{pmatrix}
\qquad\qquad\text{with}\quad
\ms{Q} \equiv \l(\vec{h}_1^{\ms{T}},\dots,\vec{h}_k^{\ms{T}}\r)\;.
\ee
Here $\vec h_j$ are an arbitrary sequence of charge vectors, $d_j=0$ if the corresponding $\vec h_j$ refers to a hyperplane of type $H_F$, $d_j = \epsilon$ if $\vec h_j$ refers to a hyperplane of type $H_V$, while $d_j = - \epsilon_a$ for a hyperplane of type $H_A^a$.

The JK-residue depends on a choice of charge vector $\vec\eta$, which plays the role of a regulator \cite{Benini:2013xpa}. When the number of hyperplanes intersecting at a point is exactly $k$ (and they are linearly independent), the singular point is called \emph{non-degenerate}. In this case the point contributes to the residue only if $\vec\eta$ is in the cone generated by the charge vectors of the hyperplanes, namely if
\be
\label{JK condition with beta's}
\ms{Q} \begin{pmatrix} \beta_1 \\ \vdots \\ \beta_k \end{pmatrix} = \vec{\eta}^{\ms{T}} \qquad\qquad\text{for some $\beta_j>0$} \;.
\ee
More generally,%
\footnote{Given a completely generic hyperplane arrangement, we do not expect more than $k$ hyperplanes to meet at a point. In the case of the elliptic genus, though, there are constraints on the fugacities: for instance because of a superpotential, or because there is no flavour fugacity associated to vector multiplets. Hence, the hyperplane arrangement associated to pole singularities of the one-loop determinant is in general degenerate.}
the number $s$ of hyperplanes through a point is larger than $k$ and the singularity is called \emph{degenerate}. In this case, computing the JK residue is more complicated. A practical method is to deform the hyperplane arrangement by adding small generic constants---not related to physical fugacities---to the arguments of the functions $\theta_1$. This ``explodes'' the degenerate singularity into $\binom{s}{k}$ non-degenerate ones. At each of the new non-degenerate singular points we compute the JK-residue, and then we sum up the various contributions. Finally, we remove the deformation in a continuous way. We analyse this method carefully in Appendix~\ref{app:desing}, reaching the explicit formula (\ref{eq:res_des_non_zero}).

We remark that, in general, the sum of JK-residues on the $u$-torus $T^{2k}$ does not depend on the choice of $\vec\eta$. In our case this would be true if we kept the multiplet $P$ throughout the computation. If, instead, we want to neglect $P$, we should make a special choice of $\vec\eta$ such that the would-be poles from $P$ would not be picked up. One can check that $\vec\eta = (1, \dots, 1)$ is such a good choice.

Let us determine the positions of poles that can have a non-vanishing JK-residue. As explained in Appendix \ref{app: technical details}, if the matrix $\ms{Q}$ solves \eqref{JK condition with beta's}, then it can be put in the form
\be
\label{eq:JKmat}
\ms{Q} =
\begin{pmatrix}
    1 & -1 & * & * & \dots & * \\
    0 &  1 & * & * & \dots & * \\
    0 &  0 & 1 & * & \dots & * \\
    \vdots & \vdots & \vdots & \vdots & \ddots & \vdots \\
    0 &  0 & 0 & 0 & \dots & 1
\end{pmatrix}
\ee
up to Weyl permutations (\ie{} up to permutations of the $u_j$'s), where each $*$ can be either $0$ or $-1$, in such a way that every column is a charge vector $\vec{h}_j$. From \eqref{eq:JKmat} we read off that the first hyperplane is of type $H_{F}$, while the other ones are either of type $H_{V;ij}$ or of the type $H_{A;ij}^a$ with $i>j$. It follows that a singular point $\{u_j\}$ can be constructed as a tree diagram with $k$ nodes. Up to Weyl permutations, the first coordinate is $u_1 = 0$. Then, each coordinate differs from one of the previous ones by either $\epsilon$ or $-\epsilon_a$.

At a singular point $\{u_j\}$, the coordinates take values on a 3d lattice
\be
\label{eq:pole_location}
U_{(l,m,n)} = (1-l)\epsilon_1 + (1-m) \epsilon_2 + (1-n) \epsilon_3 \;.
\ee
Therefore, we can alternatively represent each singular point (up to Weyl permutations) by a collection of $k$ ``boxes'' at lattice points. It turns out that only those singular points whose corresponding configuration of boxes is a \emph{plane partition} can have non-vanishing JK-residue. We prove this technical point in Appendix~\ref{app:pp}. Plane partitions are configurations such that: 1) each box sits at a different lattice point; 2) only the points $U_{ijk}$ with \mbox{$i,j,k\geq1$} can be occupied; 3) the point $U_{ijk}$ can be occupied only if all points $U_{\tilde\imath jk}$ with \mbox{$1\leq \tilde \imath<i$}, all points $U_{i\tilde\jmath k}$ with $1 \leq \tilde\jmath<j$, and all points $U_{ij\tilde k}$ with $1\leq \tilde k<k$ are also occupied. In fact, these are 3d versions of Young diagrams. For $k=1$ the only singular point (which does contribute to the JK-residue) is $u_1=0$, which is represented by a box at the origin.

\begin{table}[t]
\centering
\begin{tabular}{lrlc}
\toprule
Factor &  &Hyperplane & Order of singularity  \\
\midrule
$\jt{u_i}$ & $H_F$:  &$u_i = 0$ & $+1$ \\
$\jt{u_{ij} + \epsilon_a}$& $H_A^{(a)}$:  & $u_i = u_j - \epsilon_a$ &  $+1$ \\
$\jt{u_{ij} - \epsilon}$&  $H_V$: & $u_i = u_j + \epsilon$ &  $+1$ \\
$\jt{u_i - \epsilon}$ & $Z_F$: & $u_i = \epsilon$ & $-1$ \\
$\jt{u_{ij}}$ & $Z_V$: & $u_i = u_j$ & $-1$ \\
$\jt{u_{ij} - \epsilon_{ab}}$& $Z_A^{(ab)}$: & $u_i = u_j + \epsilon_{ab}$ & $-1$ \\
\bottomrule
\end{tabular}
\caption{Contributions to the order of singularity from the integrand in \eqref{elliptic genus N=1}.
\label{tab: sing}}
\end{table}

To each singular point we can assign an order of the singularity. Each singular hyperplane through the point contributes $+1$ to the singularity order, while each vanishing hyperplane through the point---coming from a zero of a function $\theta_1$ in the numerator---contributes $-1$. We list the possible contributions in Table~\ref{tab: sing}. A necessary condition such that a singular point has non-vanishing JK-residue is that the order of the singularity is $k$ or larger. If the singular point is non-degenerate, this simply follows from the fact that the JK-residue is an iterated residue in $\bC^k$. If the singular point is degenerate, we resolve it into $\binom{s}{k}$ non-degenerate singularities and then the statement follows from the analysis of Appendix \ref{app: technical details}. In Figure~\ref{fig:ppp} we give some examples of counting of the order.

\begin{figure}[h]
\centering
     \begin{subfigure}[t]{.49\textwidth}
      \centering
      \includegraphics[scale=.7]{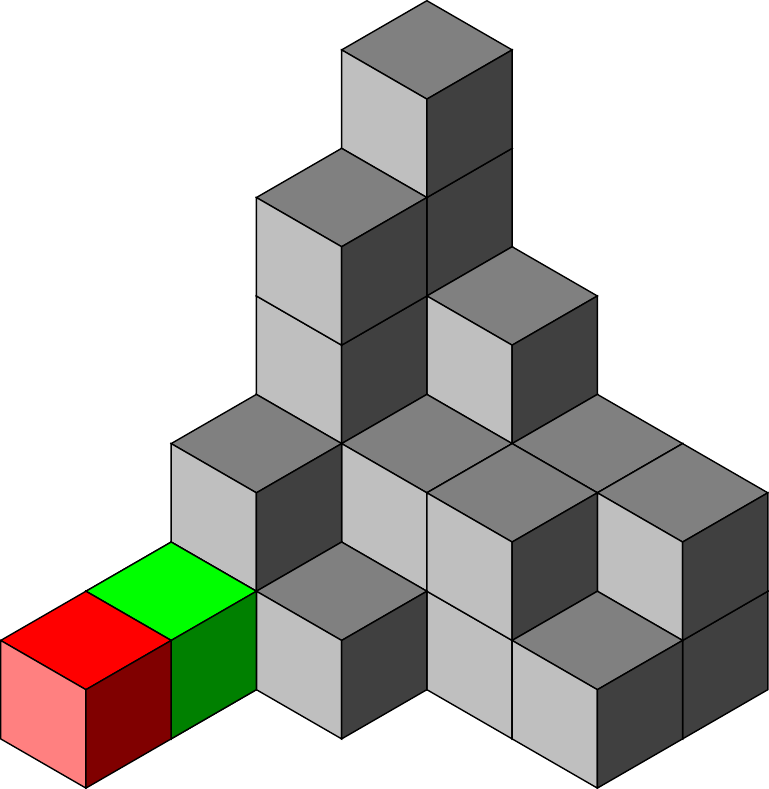}
      \caption{Adding a box along an \emph{edge.}}
     \end{subfigure}
     \hfill
     \begin{subfigure}[t]{.49\textwidth}
      \centering
      \includegraphics[scale=.7]{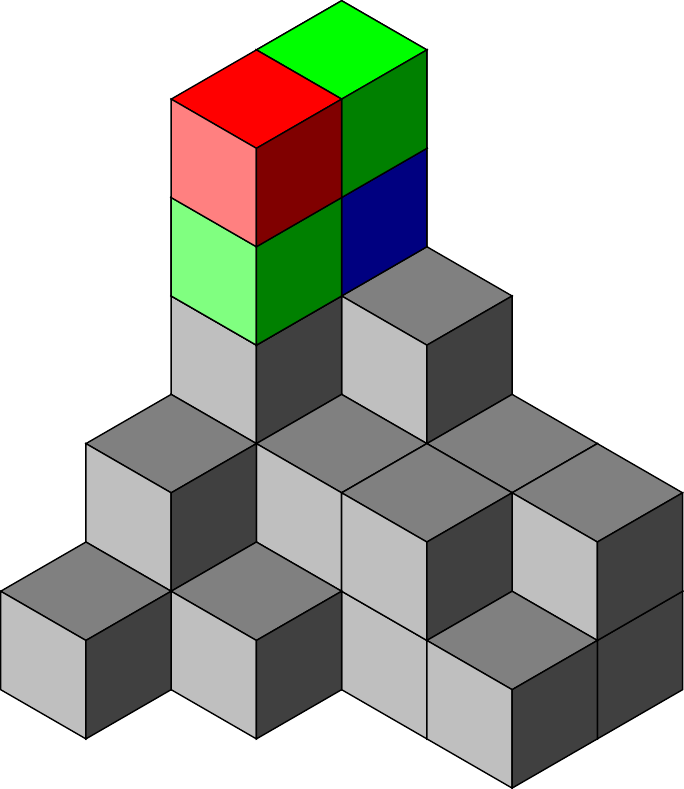}
      \caption{Adding a box to a \emph{face.}}
     \end{subfigure}
     \\[10mm]
    \begin{subfigure}[t]{.49\textwidth}
      \centering
      \includegraphics[scale=.7]{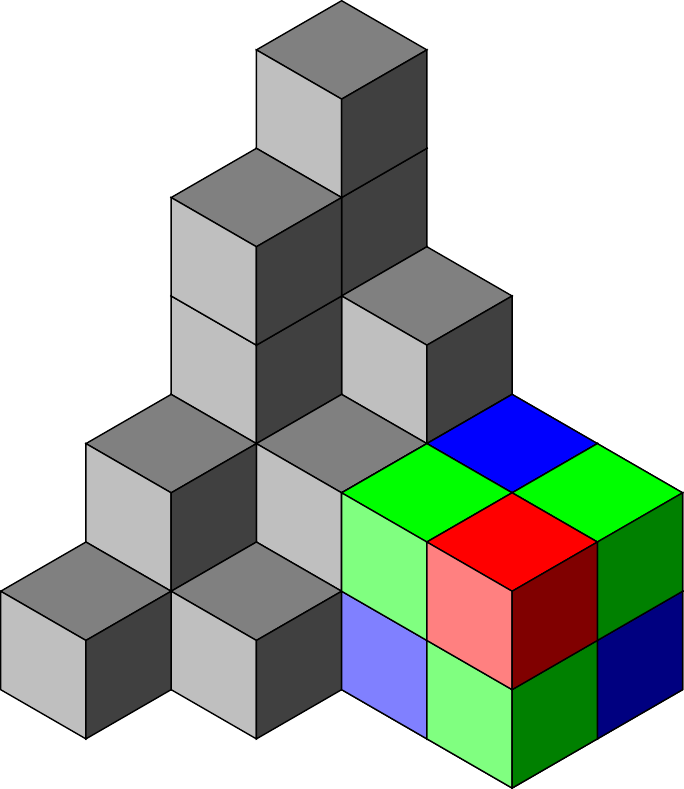}
      \caption{Adding a box to the \emph{bulk.}}
     \end{subfigure}
     \hfill
     \begin{subfigure}[t]{.49\textwidth}
      \centering
      \includegraphics[scale=.7]{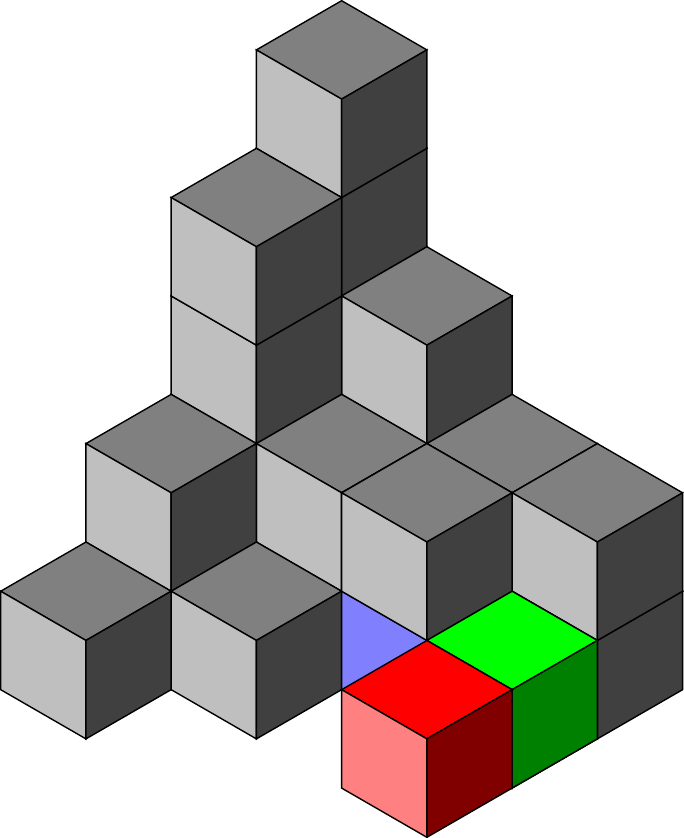}
      \caption{Adding a box such that the new arrangement is not a plane partition.}
     \end{subfigure}
\caption{Several ways to add the $(k+1)$\textsuperscript{th} box (the red one) given an arrangement of $k$ boxes. At the same time we add an integral over $u_{k+1}$. We coloured in \emph{green} those boxes whose position differs, from that of the red one, by $\epsilon_a$;  in \emph{blue} those boxes whose position differs by $\epsilon_{ab}$. From Table~\ref{tab: sing} we see that a green box increases the singularity order of the integrand by $1$, while a blue box decreases it by $1$. In case (a) we increase the order by $1$, therefore the pole contributes. In case (b) we increase the order by $2-1=1$, therefore the pole contributes. In case (c) we increase the order by $3+1-3=1$, therefore the pole contributes. In case (d) there is no change in the order of the singularity, therefore the pole does not contribute.
\label{fig:ppp}}
\end{figure}

The elliptic genus (\ref{elliptic genus N=1}) reduces to a sum of residues at those singular points that are picked up by the JK contour prescription:
\be
\label{Z^1_k as sum over partitions}
Z^{(1)}_k = \sum_{|\pi|=k}Z^{(1)}_\pi \;,
\ee
where the sum is over plane partitions with $k$ boxes. Each plane partition encodes the position of a pole. For fixed plane partition, each box at position $\vec{l}\equiv(l,m,n)$ specifies the value of one of the coordinates, $u_i = U_{(l,m,n)}$ according to (\ref{eq:pole_location}), and the order of the coordinates is not important because of the residual Weyl permutation gauge symmetry. The summands in (\ref{Z^1_k as sum over partitions}) are
\begin{multline}
\label{eq:zmid}
Z^{(1)}_\pi = \jt{\epsilon} \, \l[ - \frac{ \jt{\epsilon_{12}} \jt{\epsilon_{13}} \jt{\epsilon_{23}} }{ \jt{\epsilon_1} \jt{\epsilon_2} \jt{\epsilon_3} \jt{\epsilon} } \r]^{|\pi|}
\prod_{\vec{l} \,\in\, \pi\setminus (1,1,1)} \f{\jt{U_{\vec{l}}-\epsilon}}{\jt{U_{\vec{l}}}} \times {} \\
{} \times \sideset{}{'} \prod_{\substack{\vec{l} \,,\, \vec{l}' \,\in\, \pi \\ \vec{l} \,\neq\, \vec{l}'}}
\frac{ \jt{U_{\vec{l},\vec{l}'}} \jt{U_{\vec{l},\vec{l}'} - \epsilon_{12}} \jt{U_{\vec{l},\vec{l}'} - \epsilon_{13}} \jt{U_{\vec{l},\vec{l}'} - \epsilon_{23}} }{ \jt{U_{\vec{l},\vec{l}'}+\epsilon_1} \jt{U_{\vec{l},\vec{l}'}+\epsilon_2} \jt{U_{\vec{l},\vec{l}'}+\epsilon_3} \jt{U_{\vec{l},\vec{l}'}-\epsilon} } \;.
\end{multline}
where $U_{\vec{l},\vec{l}'}\equiv U_{\vec{l}}-U_{\vec{l}'}$. The first product is over all boxes of the plane partition, but the one located at the origin $(1,1,1)$. The second product is over all ordered pairs of boxes in the plane partition; prime means that vanishing factors, both in the numerator and denominator, are excluded from the product (as explained in Appendix~\ref{app:desing}). Many cancellations occur and the product can be recast in the form
\be
\label{eq:zfin}
Z_\pi^{(1)} = (-1)^{|\pi|} \, \frac{N_\pi^{(1)}}{D_\pi^{(1)}} \;,
\ee
where
\begin{align}
N_\pi^{(1)} &= \prod_{(r,s,t) \,\in\, \pi} \Bigg\{ \jt{r\epsilon_1 + s\epsilon_2 + \big( t-h^{xy}_{1,1} \big) \epsilon_3} \times {} \nonumber \\ \label{eq:zfinN}
& \hspace{1cm} \times \prod_{t'=1}^{h^{xy}_{1,1}} \bigg[ \jt{ \big( r-h^{yz}_{s,t'} \big) \epsilon_1 + \big( 1+h^{xz}_{r,t}-s \big) \epsilon_2 + (1+t-t') \epsilon_3} \times {} \\
& \hspace{2.3cm} \times \jt{ \big( 1+h^{yz}_{s,t'}-r \big) \epsilon_1 + \big( s-h^{xz}_{r,t} \big) \epsilon_2 + (1+t'-t ) \epsilon_3} \bigg] \Bigg\} \nonumber
\end{align}
and
\begin{align}
D_\pi^{(1)} &= \prod_{(r,s,t) \,\in\, \pi}\Bigg\{ \jt{(1-r) \epsilon_1 + (1-s)\epsilon_2 + \big( 1+h^{xy}_{1,1}-t \big) \epsilon_3} \times {} \nonumber \\ \label{eq:zfinD}
& \hspace{1cm} \times \prod_{t'=1}^{h^{xy}_{1,1}} \bigg[ \jt{ \big( r-h^{yz}_{s,t'}\big) \epsilon_1 + \big( 1+h^{xz}_{r,t}-s \big) \epsilon_2 + (t-t')\epsilon_3} \times {} \\
& \hspace{2.3cm} \times \jt{\big( 1+h^{yz}_{s,t'}-r \big) \epsilon_1 + \big( s-h^{xz}_{r,t} \big) \epsilon_2 + (t'-t )\epsilon_3} \bigg] \Bigg\} \;. \nonumber
\end{align}
Each product is over the boxes of the plane partition $\pi$. Then $h^{xy}_{r,s}$ is the depth of the pile of boxes laying at $(r,s,*)$; $h^{xz}_{r,t}$ is the height of the column of boxes at $(r,*,t)$; and $h^{yz}_{s,t}$ is the length of the row of boxes laying at $(*,s,t)$. In fact, \eqref{eq:zfin}--\eqref{eq:zfinD} are the elliptic Abelian version of similar equations in Section~4.1 of \cite{Szabo:2015wua}.

Surprisingly, we observe that for $\epsilon \in \bZ$ the expression $Z_\pi^{(1)}$ in (\ref{eq:zfin}) simplifies: as a matter of fact we find
\be
\label{Z_pi^(1)}
Z_\pi^{(1)} = (-1)^{k\epsilon} \;.
\ee
The dependence on $\epsilon$ is dictated by the 't~Hooft anomaly (\ref{'t Hooft anomaly}). There is no other dependence on $\epsilon_a$ nor on $\tau$. This implies that, up to a sign, $Z_k^{(1)}$ equals the integer number of plane partitions with $k$ boxes. It is then convenient to define a ``grand canonical'' elliptic genus, function of a new fugacity $v$, by resumming all contributions from the sectors at fixed $k$:
\be
\label{eq:grcan_ellgen}
Z^{(1)}(v) \,\equiv\, 1 + \sum_{k=1}^{\infty}Z^{(1)}_k v^k \;.
\ee
Up to a sign, this is the generating function of the number of plane partitions, namely the MacMahon function:
\be
\label{Z Abelian elliptic}
Z^{(1)}(v) = \Phi\big( (-1)^\epsilon\, v \big) \;,
\ee
where
\be
\Phi(v) \,\equiv\, \prod_{k=1}^\infty \frac1{(1-v^k)^k} = \pe{v}{\frac v{(1-v)^2}}
\ee
is the MacMahon function and $\mathtt{PE}$ is the plethystic exponential operator (see Appendix \ref{app: plethystic exp}).

\subsection{Dimensional Reductions}

We can consider dimensional reductions of the system. Reducing on a circle, we obtain the Witten index of an $\cN=4$ SUSY quantum mechanics. This case, known as \emph{trigonometric} or \emph{motivic}, has been studied in \cite{Nekrasov:2009jap}. It can be obtained from the elliptic case in the limit $p \to 0$, where $p=e ^{2\pi i \tau}$. By a further reduction on a second circle, we obtain a SUSY matrix integral with 4 supercharges. This case, known as \emph{rational}, has been studied in \cite{Szabo:2015wua}. It can be obtained from the trigonometric case in the limit $\beta \to 0$, where $\beta$ is the radius of the circle used to compute the Witten index in the path integral formulation.

It is important to notice that in the trigonometric and rational cases, corresponding to field theories in 1d and 0d respectively, there is no anomaly constraint and one can take generic real values for the parameter descending from $\epsilon$. This means that, in order to have access to all values of the parameters, we should apply the two limits to the integrand in (\ref{elliptic genus N=1}) and then recompute the contour integral.

Given a quantity $X$ in the elliptic case, we use the notation $\wt X$ for the corresponding quantity in the trigonometric case and $\wb X$ in the rational case. We also use $\bullup X$ to refer to the three cases at the same time.

\subsubsection{Trigonometric limit}

To obtain the trigonometric limit, we use that $\jt{z} \to 2 p^{1/8} \sin(\pi z)$ as $p \to 0$. We express the result in terms of new variables
\be
\label{def parameters QM}
q_a = e^{2\pi i \epsilon_a} \;,\qquad q = e^{2\pi i \epsilon} \;,\qquad x_i = e^{2\pi i u_i} \;,\qquad p=e^{2\pi i \tau} \;,
\ee
with $q_1q_2q_3 = q$. We find the integral expression for the Witten index of the $\cN=4$ SUSY quantum mechanics corresponding to the quiver in Figure~\ref{fig: quiver}:
\begin{multline}
\label{eq:int_rat}
\wt{Z}^{(1)}_k(q_a) = \frac1{k!} \l[-q^{\f{1}{2}} \f{ (1-q_1q_2)(1-q_1q_3)(1-q_2q_3) }{ (1-q_1)(1-q_2)(1-q_3)(1-q) } \r]^k \int_\text{JK} \prod_{i=1}^k \f{\d x_i}{x_i} \prod_{i=1}^k \f{1-q^{-1}x_i}{1-x_i} \times {} \\
{} \times \prod_{\substack{i,j=1 \\ i\neq j}}^k q\f{ (1-x_{ij})(1-q_1^{-1}q_2^{-1}x_{ij})(1-q_1^{-1}q_3^{-1}x_{ij})(1-q_2^{-1}q_3^{-1}x_{ij}) }{ (1-q_1x_{ij})(1-q_2x_{ij})(1-q_3x_{ij})(1-q^{-1}x_{ij}) } \;.
\end{multline}
Since there are no anomalies this time, the value of $\epsilon$ is unconstrained.
The Witten index of SUSY quantum mechanics can jump when flat directions open up at infinity in field space. From the point of view of the 7D theory on the D6-brane, or DT invariants of $\bC^3$, this is the wall crossing phenomenon. In the quantum mechanics, the parameter we vary is the Fayet-Iliopoulos (FI) term and it corresponds to the stability parameter in DT theory. The integral in \eqref{eq:int_rat} is a contour integral in $(\bC^*)^k$, and in general it includes boundary components. However, choosing the auxiliary parameter $\vec\eta$ parallel to the FI parameter guarantees that the JK contour has no boundary components \cite{Hwang:2014uwa, Cordova:2014oxa, Hori:2014tda} (see also \cite{Benini:2015noa, Benini:2016hjo}). The chamber with non-trivial DT invariants corresponds to $\vec\eta = (1, \dots, 1)$.

The result can be expressed as before:
\be
\label{eq:ztrigfin}
\wt Z^{(1)}_k= \sum_{|\pi|=k} \wt{Z}^{(1)}_\pi \;, \qquad\qquad\qquad \wt{Z}^{(1)}_\pi= (-1)^{|\pi|} \, \frac{\wt{N}^{(1)}_\pi}{\wt{D}^{(1)}_\pi} \;,
\ee
where
\begin{align}
\wt N_\pi^{(1)} &= \prod_{(r,s,t) \,\in\, \pi}\Bigg\{ \af{q_1^r \, q_2^s \, q_3^{t-h^{xy}_{1,1}}} \times {} \nonumber \\
& \hspace{1cm} {} \times \prod_{t'=1}^{h^{xy}_{1,1}} \bigg[ \af{q_1^{r-h^{yz}_{s,t'}} q_2^{1+h^{xz}_{r,t}-s} q_3^{1+t-t'}} \, \af{q_1^{1+h^{yz}_{s,t'}-r} q_2^{s-h^{xz}_{r,t} } q_3^{1+t'-t }} \bigg] \Bigg\}
\label{eq:ztrigfinN} \\
\wt{D}_\pi^{(1)} &= \prod_{(r,s,t) \,\in\, \pi} \Bigg\{ \af{q_1^{1-r} \, q_2^{(1-s)} \, q_3^{(1+h^{xy}_{1,1}-t}} \times {} \nonumber \\
& \hspace{1cm} {} \times \prod_{t'=1}^{h^{xy}_{1,1}} \bigg[ \af{q_1^{r-h^{yz}_{s,t'}} q_2^{1+h^{xz}_{r,t}-s} q_3^{t-t'}} \, \af{q_1^{1+h^{yz}_{s,t'}-r} q_2^{s-h^{xz}_{r,t}} q_3^{t'-t }}\bigg] \Bigg\} \;.
\label{eq:ztrigfinD}
\end{align}
The notation is the same as in \eqref{eq:zfinN} and \eqref{eq:zfinD}. We defined the function
\be
\af{x} = x^{\f{1}{2}} - x^{-\f{1}{2}} \;,
\ee
in other words $\af{e^{2\pi i z}} = 2i \sin(\pi z)$. Notice that \eqref{eq:ztrigfinN} and \eqref{eq:ztrigfinD} are simply obtained from \eqref{eq:zfinN} and \eqref{eq:zfinD} by substituting $\jt{z} \mapsto \sin(\pi z)$, because the extra powers of $p$ cancel out.

\subsubsection{Rational limit}

To obtain the rational limit, we place the SUSY quantum mechanics on a circle of radius $\beta$ and shrink it. This can be done, starting from (\ref{def parameters QM}) and \eqref{eq:int_rat}, by substituting $\epsilon_a \mapsto \beta\epsilon_a$ and $u_i \mapsto \beta u_i$, then taking a $\beta \to 0$ limit. The result is
\begin{multline}
\wb{Z}_k^{(1)}(\epsilon_a) = \frac{1}{k!} \bigg[ \f{\epsilon_{12}\epsilon_{13}\epsilon_{23}}{\epsilon_1\epsilon_2\epsilon_3\epsilon} \bigg]^k \int_\text{JK} \prod_{i=1}^k \d u_i \prod_{i=1}^k\f{u_i-\epsilon}{u_i} \times {} \\
{} \times \prod_{\substack{i,j=1 \\ i\neq j}} \f{u_{ij}(u_{ij} - \epsilon_{12})(u_{ij} - \epsilon_{13})(u_{ij}-\epsilon_{23})}{(u_{ij} + \epsilon_{1})(u_{ij}+\epsilon_{2})(u_{ij}+\epsilon_{3})(u_{ij}-\epsilon)} \;.
\end{multline}
This expression can be cast in the same form as in previous cases:
\be
\label{eq:zratfin}
\wb{Z}^{(1)}_k = \sum_{|\pi|=k} \wb{Z}^{(1)}_\pi \;, \qquad\qquad\qquad \wb{Z}^{(1)}_\pi= (-1)^{|\pi|}\f{\wb{N}^{(1)}_\pi}{\wb{D}^{(1)}_\pi} \;,
\ee
with
\begin{align}
\wb N_\pi^{(1)} &= \prod_{(r,s,t) \,\in\,\pi} \Bigg\{ \Big( r\epsilon_1 + s\epsilon_2 + \big( t-h^{xy}_{1,1} \big) \epsilon_3 \Big) \times {} \nonumber \\
& \hspace{1cm} \times \prod_{t'=1}^{h^{xy}_{1,1}} \bigg[ \Big( \l(r-h^{yz}_{s,t'}\r)\epsilon_1 + \l(1+h^{xz}_{r,t}-s\r)\epsilon_2 + (1+t-t')\epsilon_3 \Big) \times {}
\label{eq:zratfinN} \\
& \hspace{1.5cm} \times \Big( \l(1+h^{yz}_{s,t'}-r\r)\epsilon_1 + \l(s-h^{xz}_{r,t} \r)\epsilon_2 + (1+t'-t )\epsilon_3 \Big) \bigg] \Bigg\} \nonumber \\
\wb D_\pi^{(1)} &= \prod_{(r,s,t) \,\in\, \pi} \Bigg\{ \Big( (1-r)\epsilon_1 + (1-s)\epsilon_2 + \l(1+h^{xy}_{1,1}-t\r) \epsilon_3 \Big) \times {} \nonumber \\
& \hspace{1cm} \times \prod_{t'=1}^{h^{xy}_{1,1}} \bigg[ \Big( \l(r-h^{yz}_{s,t'}\r)\epsilon_1 + \l(1+h^{xz}_{r,t}-s\r)\epsilon_2 + (t-t')\epsilon_3 \Big) \times {}
\label{eq:zratfinD} \\
& \hspace{1.5cm} \times \Big( \l(1+h^{yz}_{s,t'}-r\r)\epsilon_1 + \l(s-h^{xz}_{r,t} \r)\epsilon_2 + (t'-t )\epsilon_3 \Big) \bigg] \Bigg\} \;. \nonumber
\end{align}
Once again, \eqref{eq:zratfinN} and \eqref{eq:zratfinD} are obtained from \eqref{eq:zfinN} and \eqref{eq:zfinD} by substituting $\jt{z} \mapsto z$.

\subsection{The plethystic ans\"atze}
\label{sec: plethystic}

As we observed in (\ref{Z_pi^(1)})--(\ref{Z Abelian elliptic}), the elliptic Abelian DT invariants are very simple and count the number of plane partitions. This is because the dependence of the elliptic genera on $\epsilon \in \bZ$ is fixed by the anomaly, and there is no dependence on $\tau$. The latter is a general property of gapped systems (see \eg{} \cite{Benini:2013xpa} for other examples) due to the fact that the elliptic genus of a gapped vacuum does not depend on $\tau$.

By dimensional reduction, this implies that also the trigonometric and rational DT invariants, evaluated at $\epsilon=0$, are captured by MacMahon's function. Defining a grand canonical partition function
\be
\accentset{\bullet}{Z}^{(1)}(v) \,\equiv\, 1 + \sum_{k=1}^\infty \accentset{\bullet}{Z}^{(1)}_k \, v^k
\ee
both in the elliptic, trigonometric and rational case, we find that they are all equal to the MacMahon function:
\be
Z^{(1)}(v) \big|_{\epsilon=0} = \wt Z^{(1)}(v) \big|_{\epsilon=0} = \wb Z^{(1)}(v) \big|_{\epsilon=0} = \Phi(v) \;.
\ee
In the trigonometric and rational case, it is natural to ask whether a similar plethystic expression holds also when $\epsilon\neq 0$ (since there is no constraint on $\epsilon$). It is clear that such an expression cannot be derived from the elliptic case.

It has been proved in \cite{MNOP06a, MNOP06b} that in the rational case the grand canonical partition function is simply
\be
\label{eq:ratgcab}
\wb{Z}^{(1)} = \Phi(v)^{-\f{\epsilon_{12}\epsilon_{13}\epsilon_{23}}{\epsilon_1\epsilon_2\epsilon_3}} = \pe{v}{ -\f{ \epsilon_{12} \epsilon_{13} \epsilon_{23}}{ \epsilon_1 \epsilon_2 \epsilon_3} \, \f{v}{(1-v)^2}} \;.
\ee
Notice that in this formula the plethystic variable is just $v$ (not $\epsilon_a$).
In the trigonometric case, the following plethystic expression was conjectured by Nekrasov \cite{Nekrasov:2009jap}:
\be
\label{eq:nekans}
\wt{Z}^{(1)} = \pe{v;\vec{q}}{-\f{(1-q_1q_2)(1-q_1q_3)(1-q_2q_3)}{(1-q_1)(1-q_2)(1-q_3)} \, \f{v }{ q^{\f{1}{2}} (1-vq^{-\f{1}{2}})(1-vq^{\f{1}{2}})} } \;.
\ee
We have verified that this expression reproduces (\ref{eq:ztrigfin}) up to $k=12$.

\section{Non-Abelian case}
\label{sec: non-Abelian case}

In this section we extend the computation of the elliptic genus to quiver theories as in Figure~\ref{fig: quiver} with $N>1$. The flavour symmetry of such theories contains an $\mr{SU}(N)$ factor, as summarised in Table~\ref{tab: holo}. We add fugacities $z_\alpha$ along the Cartan generators of $\mr{SU}(N)$, with the constraint $\sum_{\alpha=1}^N z_\alpha = 0$. The elliptic genus is computed by the following contour integral \cite{Benini:2013nda, Benini:2013xpa}, that generalises (\ref{elliptic genus N=1}):
\begin{align}
\label{elliptic genus generic N}
& Z_k^{(N)}(z_\alpha,\epsilon_a,\tau) = \f{1}{k!} \left[  \frac{ 2\pi\eta^3(q) \, \jt{\epsilon_{12}} \jt{\epsilon_{13}} \jt{\epsilon_{23}}
  }{
  \jt{\epsilon_1} \jt{\epsilon_2} \jt{\epsilon_3} \jt{\epsilon} } \right]^k
   \int_\text{JK} \prod_{i=1}^k \d u_i\times {} \\
& \times
   \prod_{i=1}^k\prod_{\alpha=1}^N \frac{ \jt{u_i + z_\alpha -\epsilon}}{\jt{u_i + z_\alpha}}
   \prod_{\substack{i,j=1\\i\neq j}}^k \frac{
   \jt{u_{ij}}
   \jt{u_{ij}-\epsilon_{12}}
   \jt{u_{ij}-\epsilon_{13}}
   \jt{u_{ij}-\epsilon_{23}}
   }{
   \jt{u_{ij}+\epsilon_{1}}
   \jt{u_{ij}+\epsilon_{2}}
   \jt{u_{ij}+\epsilon_{3}}
   \jt{u_{ij}-\epsilon}
   }. \nn
\end{align}
Because of the gauge-R-symmetry anomaly, the elliptic genus is well-defined only for
\be
\label{anomaly constraint non-Abelian}
\epsilon \in \frac1N \bZ \;.
\ee
This ensures that the integrand be doubly periodic under $u_i \to u_i + a + b\tau$ with $a,b \in \bZ$. Besides, the R-symmetry 't~Hooft anomaly dictates
\be
\label{quasi-periodicity}
Z_k^{(N)} \,\to\, (-1)^{Nk} Z_k^{(N)}
\ee
when we shift one of $\epsilon_a \to \epsilon_a +1$.

We evaluate the contour integral in the same way as we did in Section~\ref{sec: Abelian case}---with technical details collected in Appendix \ref{app: technical details}---but keeping into account the fugacities for the flavour group. When $N>1$, the charge matrix $\ms{Q}$ is block diagonal, and the blocks (one for each flavour) look like (\ref{eq:JKmat}). The poles live on the union of $N$ different lattices
\be
U_{\alpha,(l,m,n)} \equiv -z_\alpha + U_{(l,m,n)} = -z_\alpha + (1-l)\epsilon_1 + (1-m)\epsilon_2 + (1-n)\epsilon_3 \;.
\ee
Representing poles by arrangements of boxes on the collection of lattices, it turns out that the poles contributing to the JK residue are those represented by $N$ distinct plane partitions labelled by $\alpha$. Such type of arrangement is known as a \emph{coloured} plane partition (see Appendix~\ref{App:pp}). We denote a coloured plane partition as $\vec{\pi} = (\pi_1,\dts, \pi_N)$. The partition function is then a sum of residues
\be
Z_k^{(N)} = \sum_{|\vec\pi| = k} Z_{\vec\pi}^{(N)}
\ee
at those poles classified by coloured plane partitions.

In order to compute the residue at a pole represented by a coloured plane partition $\vec\pi$, we observe that there are no factors in the denominator involving more than one $z_\alpha$. It follows that the residue can be written as
\begin{align}
\label{eq:znmid}
& Z^{(N)}_{\vec{\pi}}= \prod_{\pi\in\vec{\pi}}Z^{(1)}_\pi \times \prod_{\substack{\pi_\alpha, \pi_\beta \in \, \vec{\pi} \\ \alpha \neq \beta}}
  \Bigg[ \prod_{\vec{l} \in\pi_\alpha} \f{\jt{U_{\vec{l}} - z_{\alpha\beta} - \epsilon}}{\jt{U_{\vec{l}} - z_{\alpha\beta}}} \times {} \\
& \times \prod_{\substack{\vec{l} \in\pi_\alpha \\ \vec{l}' \in\pi_\beta}} \frac{
  \jt{U_{\vec{l},\vec{l}'} {-} z_{\alpha\beta}}
  \jt{U_{\vec{l},\vec{l}'} {-} z_{\alpha\beta} {-} \epsilon_{12}}
  \jt{U_{\vec{l},\vec{l}'} {-} z_{\alpha\beta} {-} \epsilon_{13}}
  \jt{U_{\vec{l},\vec{l}'} {-} z_{\alpha\beta} {-} \epsilon_{23}}
  }{
  \jt{U_{\vec{l},\vec{l}'} {-} z_{\alpha\beta} {+} \epsilon_{1}}
  \jt{U_{\vec{l},\vec{l}'} {-} z_{\alpha\beta} {+} \epsilon_{2}}
  \jt{U_{\vec{l},\vec{l}'} {-} z_{\alpha\beta} {+} \epsilon_{3}}
  \jt{U_{\vec{l},\vec{l}'} {-} z_{\alpha\beta} {-} \epsilon}
  } \Bigg] \,. \nn
\end{align}
Here $Z_\pi^{(1)}$ is the expression (\ref{eq:zmid}) from the Abelian case, while $z_{\alpha\beta} = z_\alpha - z_\beta$. We have indicated by $\vec{l}\equiv (l,m,n)$ the positions of the boxes in a plane partition, then $U_{\vec l} \equiv U_{(l,m,n)}$ and $U_{\vec{l},\vec{l}'} \equiv U_{\vec{l}}-U_{\vec{l}'}$. We stress that $U_{\vec l}$ does not depend on $z_\alpha$, as this is different from $U_{\alpha, (l,m,n)}$.

Also in this case, several cancellations occur in evaluating \eqref{eq:znmid} and it is possible to recast the result in a form similar to \eqref{eq:zfin}--\eqref{eq:zfinD}. We find:
\be
\label{eq:znfin}
Z_{\vec{\pi}}^{(N)} = (-1)^{N|\vec{\pi}|} \, \prod_{\alpha,\beta=1}^N
 \f{N_{\vec{\pi},\alpha\beta}^{(N)}(z_{\alpha\beta})}{D_{\vec{\pi},\alpha\beta}^{(N)}(z_{\alpha\beta})} \;,
\ee
with:
\begin{align}
N_{\vec{\pi},\alpha\beta}^{(N)}(z) &= \prod_{(r,s,t)\in\pi_{\alpha}} \Bigg\{
  \jt{ z+ r\epsilon_1 + s\epsilon_2 + \big( t-h^{xy;\beta}_{1,1} \big)\epsilon_3} \times {} \label{eq:znfinN} \\ 
  & \hspace{1cm} \times
  \prod_{t'=1}^{h^{xy;\beta}_{1,1}} \bigg[
  \jt{ z+ \big( r-h^{yz;\beta}_{s,t'} \big) \epsilon_1 + \big( 1+h^{xz;\alpha}_{r,t}-s \big)\epsilon_2 + (1+t-t')\epsilon_3} \times \nn \\
  & \hspace{1.4cm} \times
  \jt{ -z+\big( 1+h^{yz;\beta}_{s,t'}-r \big) \epsilon_1 + \big( s-h^{xz;\alpha}_{r,t} \big) \epsilon_2 + (1+t'-t ) \epsilon_3} \bigg]\Bigg\} \;, \nn
\end{align}
\begin{align}
D_{\vec{\pi},\alpha\beta}^{(N)}(z) &= \prod_{(r,s,t)\in\pi_\alpha} \Bigg\{
  \jt{ -z+(1-r) \epsilon_1 + (1-s)\epsilon_2 + \big( 1+h^{xy;\beta}_{1,1}-t \big) \epsilon_3} \times \label{eq:znfinD} \\ 
  &\hspace{1cm} \times
  \prod_{t'=1}^{h^{xy;\beta}_{1,1}} \bigg[
  \jt{z+ \big( r-h^{yz;\beta}_{s,t'} \big) \epsilon_1 + \big(1+h^{xz;\alpha}_{r,t}-s \big) \epsilon_2 + (t-t')\epsilon_3 } \times \nn \\
  &\hspace{1.5cm} \times
  \jt{-z+\big( 1+h^{yz;\beta}_{s,t'}-r \big) \epsilon_1 + \big( s-h^{xz;\alpha}_{r,t} \big) \epsilon_2 + (t'-t )\epsilon_3 } \bigg] \Bigg\} \;. \nn
\end{align}
Notice that now the function $h$ has an index $\alpha$ that clarifies which plane partition in the coloured set it refers to. These expressions are the elliptic version of similar equations in \cite{Szabo:2015wua}, where the rational case was analysed.

The dimensional reduction of these formulas to the trigonometric case is the following:
\be
\label{eq:zntrigfin}
\wt{Z}_{\vec{\pi}}^{(N)}= (-1)^{N|\vec{\pi}|} \, \prod_{\alpha,\beta=1}^N
\frac{ \wt{N}_{\vec{\pi},\alpha\beta}^{(N)}(a_{\alpha\beta}) }{ \wt{D}_{\vec{\pi},\alpha\beta}^{(N)}(a_{\alpha\beta})} \;,
\ee
where we set $a_\alpha=e^{2\pi i z_\alpha}$, $a_{\alpha\beta} = a_\alpha/ a_\beta$ and
\begin{align}
& \wt{N}_{\vec{\pi},\alpha\beta}^{(N)}(a) = \prod_{(r,s,t)\in\pi_\alpha} \Bigg\{
  \af{ a \, q_1^r \, q_2^s \, q_3^{t-h^{xy;\beta}_{1,1}}} \times \label{eq:zntrigfinN} \\
& \hspace{1cm} \times
  \prod_{t'=1}^{h^{xy;\beta}_{1,1}} \bigg[
  \af{ a \, q_1^{r-h^{yz;\beta}_{s,t'}} q_2^{1+h^{xz;\alpha}_{r,t}-s} q_3^{1+t-t'} }
  \af{ a^{-1} \, q_1^{1+h^{yz;\beta}_{s,t'}-r} q_2^{s-h^{xz;\alpha}_{r,t} } q_3^{1+t'-t } } \bigg] \Bigg\} \;, \nn
\end{align}
\begin{align}
& \wt{D}_{\vec{\pi},\alpha\beta}^{(N)}(a) = \prod_{(r,s,t)\in\pi_\alpha} \Bigg\{
  \af{ a^{-1} \, q_1^{1-r} \, q_2^{(1-s)} \, q_3^{(1+h^{xy;\beta}_{1,1}-t}} \times \label{eq:zntrigfinD} \\
& \hspace{1cm} \times
  \prod_{t'=1}^{h^{xy;\beta}_{1,1}} \bigg[
  \af{ a \, q_1^{r-h^{yz;\beta}_{s,t'}} q_2^{1+h^{xz;\alpha}_{r,t}-s} q_3^{t-t'} }
  \af{a^{-1} \, q_1^{1+h^{yz;\beta}_{s,t'}-r} q_2^{s-h^{xz;\alpha}_{r,t}} q_3^{t'-t } } \bigg] \Bigg\} \;. \nn
\end{align}

The reduction to the rational case gives the following:
\be
\label{eq:znratfin}
\wb{Z}_{\vec{\pi}}^{(N)} = (-1)^{N|\vec{\pi}|} \, \prod_{\alpha,\beta=1}^N \frac{ \wb{N}_{\vec{\pi},\alpha\beta}^{(N)}(z_{\alpha\beta}) }{ \wb{D}_{\vec{\pi},\alpha\beta}^{(N)}(z_{\alpha\beta})} \;,
\ee
with
\begin{align}
\wb{N}_{\vec{\pi},\alpha\beta}^{(N)}(z) &= \prod_{(r,s,t)\in\pi_{\alpha}} \Bigg\{
  \Big( z+r\epsilon_1 + s\epsilon_2 + \l(t-h^{xy;\beta}_{1,1}\r)\epsilon_3 \Big) \times \label{eq:znratfinN} \\
& \hspace{1cm} \times
  \prod_{t'=1}^{h^{xy;\beta}_{1,1}} \bigg[
  \Big( z+ \l(r-h^{yz;\beta}_{s,t'} \r) \epsilon_1 + \l(1+h^{xz;\alpha}_{r,t}-s\r)\epsilon_2 + (1+t-t')\epsilon_3 \Big) \times \nn \\
& \hspace{1.3cm} \times
  \Big( -z+\l(1+h^{yz;\beta}_{s,t'}-r \r)\epsilon_1 + \l(s-h^{xz;\alpha}_{r,t} \r)\epsilon_2 + (1+t'-t )\epsilon_3 \Big) \bigg] \Bigg\}  \;, \nn
\end{align}
\begin{align}
\wb{D}_{\vec{\pi},\alpha\beta}^{(N)}(z) &= \prod_{(r,s,t)\in\pi_\alpha} \Bigg\{
  \Big( -z+(1-r)\epsilon_1 + (1-s)\epsilon_2 + \l(1+h^{xy;\beta}_{1,1}-t \r)\epsilon_3 \Big) \times \label{eq:znratfinD} \\ 
& \hspace{1cm} \times
  \prod_{t'=1}^{h^{xy;\beta}_{1,1}} \bigg[
  \Big( z+\l(r-h^{yz;\beta}_{s,t'}\r)\epsilon_1 + \l(1+h^{xz;\alpha}_{r,t}-s\r)\epsilon_2 + (t-t')\epsilon_3 \Big) \times \nn \\
& \hspace{1.5cm} \times
  \Big( -z+\l(1+h^{yz;\beta}_{s,t'}-r\r)\epsilon_1 + \l(s-h^{xz;\alpha}_{r,t} \r)\epsilon_2 + (t'-t )\epsilon_3 \Big) \bigg] \Bigg\} \;. \nn 
\end{align}
This reproduces the expressions in Section 4 of \cite{Szabo:2015wua}.

\subsection{Resummation conjectures and factorisation}

We are interested in the generating functions of non-Abelian Donaldson-Thomas invariants, namely in the ``grand canonical'' partition functions
\be
\accentset{\bullet}{Z}^{(N)}(v) = 1+ \sum_{k=1}^{\infty}\accentset{\bullet}{Z}^{(N)}_k \, v^k \;,
\ee
in the three cases---elliptic, trigonometric and rational.

As in the Abelian case, we observe that \eqref{eq:znfin}, \eqref{eq:zntrigfin} and \eqref{eq:znratfin} drastically simplify when we set $\epsilon=0$:
\be
\label{simplification epsilon=0}
Z_{\vec\pi}^{(N)} \big|_{\epsilon=0} = \wt Z_{\vec\pi}^{(N)} \big|_{\epsilon=0} = \wb Z_{\vec\pi}^{(N)} \big|_{\epsilon=0} = 1 \;.
\ee
This implies that the grand canonical partition function reduces to the $N^\text{th}$ power of MacMahon's function,
\be
Z^{(N)} \big|_{\epsilon=0} = \wt Z^{(N)} \big|_{\epsilon=0} = \wb Z^{(N)} \big|_{\epsilon=0} = \Phi(v)^N \;,
\ee
with no dependence on the flavour fugacities, nor on $\tau$ in the elliptic case.

Next, we observe that in all cases the dependence on the flavour fugacities cancels out in $\accentset\bullet{Z}^{(N)}_k$, after summing the various contributions from coloured plane partitions. We have verified this claim up to a certain order in $k$. Assuming that the cancellation persists to all orders, our task of identifying the grand canonical partition functions simplifies.

Let us start with the elliptic DT invariants. As opposed to the Abelian case, for $N>1$ (\ref{simplification epsilon=0}) and the anomalous quasi-periodicity (\ref{quasi-periodicity}) are not enough to fix the partition function, since now $\epsilon = n/N$ with $n \in \bZ$. Nevertheless, inspecting the result for various values of $N$ and $k$, we were able to propose the following formula:
\be
\label{eq:ZkN}
Z^{(N)}_k\Big|_{\epsilon = \f{n}{N}} =
  \begin{cases}
   (-1)^{nk} \; \Phi^{\left( \rule{0pt}{.6em} \gcd(n,N) \right)}_{\f{k}{N}\gcd(n,N)} & \text{if } \f{N}{\gcd(n,N)} | k \;, \\[.5em]
   0 & \text{otherwise} \;.
  \end{cases}
\ee
Here the coefficients $\Phi^{(N)}_k$, defined in Appendix~\ref{App:pp}, are those of the series expansion of $\Phi(v)^N$. Moreover recall that $\gcd(0,N)=N$. The proposal (\ref{eq:ZkN}) satisfies the anomalous quasi-periodicity (\ref{quasi-periodicity}). It is then easy to resum the series:
\be
\label{eq:Ztot}
Z^{(N)}\big|_{\epsilon=\f{n}{N}}(v) = \Phi\l((-1)^{nN} \, v^{\f{N}{\gcd(n,N)}}\r)^{\gcd(n,N)} \;.
\ee
We provide a string theory derivation of this formula in Section~\ref{sec: F-theory}. As in the Abelian case, we should expect no dependence on $\tau$ because the two-dimensional theory is gapped. The lack of dependence on the flavour fugacities is also observed in other gapped models, for instance the Grassmannians (see \eg{} \cite{Benini:2013xpa}).

In the trigonometric case, the following expression was proposed in \cite{Awata:2009dd}:%
\footnote{We have verified it up to $k=5$ and $N=5$.} 
\be
\label{eq:facttrig}
\wt{Z}^{(N)}=\pe{v,\vec{q}}{-\f{(1-q_1q_2)(1-q_1q_3)(1-q_2q_3)}{(1-q_1)(1-q_2)(1-q_3)}q^{-\f{N}{2}}\f{1-q^N}{1-q}\f{v}{(1-vq^{-\f{N}{2}})(1-vq^{\f{N}{2}})}} \;.
\ee
This reproduces Nekrasov's ansatz \eqref{eq:nekans} for $N=1$. We provide an M-theory derivation of this formula in Section~\ref{sec: M-theory}. It is possible to show that
\be
\label{eq:ZZ}
\wt{Z}^{(N)} \big|_{\epsilon=\f{n}{N}} = Z^{(N)} \big|_{\epsilon=\f{n}{N}} \;.
\end{equation}
In order to evaluate the left-hand-side some care is needed: if we set $q=e^{2\pi i\f{n}{N}}$ we find a vanishing argument in the plethystic exponential. Applying the definition \eqref{eq:pletdef}, though, we see that the terms that survive in the expansion are those for which $\f{kn}{N}\in\mb{Z}$, namely such that $\f{N}{\gcd(n,N)}|k$. We can compute those terms by substituting $n\mapsto \alpha n$ and the taking the limit $\alpha\rightarrow 1$. 

Finally, for the rational case a conjecture was already put forward in \cite{Szabo:2015wua,Nekrasov:2009jap}:
\be
\label{eq:factrat}
\wb{Z}^{(N)}(v)=\l(\wb{Z}^{(1)}(v) \r)^N = \Phi(v)^{-N\f{\epsilon_{12}\epsilon_{13}\epsilon_{23}}{\epsilon_1\epsilon_2\epsilon_3}} \;.
\end{equation}
We have verified this conjecture up to $k=8$ and $N=8$. As a check, the trigonometric expression (\ref{eq:facttrig}) reduces to (\ref{eq:factrat}) in the rational limit. It is particularly simple to see that the trigonometric expression has a well-defined $q\to1$ limit yielding $\Phi(v)^N$.

\subsection{F-theoretic interpretation of elliptic DT counting}
\label{sec: F-theory}

We can give an interpretation of the elliptic non-Abelian DT invariants (\ref{eq:ZkN}) from their realisation in type IIB string theory, or F-theory, in terms of the D1/D7 brane system.

The setup consists of $N$ D7-branes wrapping $T^2 \times \bC^3$, as well as $k$ D1-branes on the worldvolume of the D7's and wrapping $T^2$. There is a further complex plane $\bC$ orthogonal to all branes. We can introduce a complex coordinate $w$ on $T^2$, complex coordinates $x_{1,2,3}$ on $\bC^3$ and $u$ on $\bC$. The $\Omega$-background is geometrically implemented by fibering $\bC^3 \times \bC$ on $T^2$ in a non-trivial way, controlled by four complex parameters $\epsilon_{1,2,3,4}$. The fibering of complex structure that corresponds to the scheme we chose in field theory is such that each of the complex factors in the fiber is rotated by a complexified phase $e^{2\pi i \epsilon_a}$ for $a=1,2,3,4$, respectively, when we go around the B-cycle of $T^2$, while they are not rotated when we go around the A-cycle. Supersymmetry requires to impose a Calabi-Yau condition to the total geometry, $\sum_{a=1}^4 \epsilon_a = 0$. This means that we can identify $\epsilon_4 = - \epsilon = - \sum_{a=1}^3 \epsilon_a$.

The D7-branes source a non-trivial holomorphic profile for the axio-dilaton $\tau_\text{IIB}$ along the $\bC$ fiber:
\be
\tau_\text{IIB}(z) = \frac1{2\pi i} \sum_{\alpha=1}^N \log (u-u_\alpha) \;,
\ee
where $u_\alpha$ are the positions of the D7-branes on $\bC$. Such parameters are controlled by real masses associated to the $\mathrm{SU}(N)$ flavour symmetry in field theory. Going around the B-cycle, the fiber is rotated as $u \to e^{-2\pi i \epsilon}u$. Considering the case $u_\alpha=0$, the condition that the axio-dilaton be periodic up to $SL(2,\bZ)$ transformations imposes the constraint
\be
N \epsilon \in \bZ \;.
\ee
This reproduces the anomaly constraint (\ref{anomaly constraint non-Abelian}) in field theory, and forces us to set $\epsilon = n/N$ with $n\in\bZ$.

\begin{figure}[h]
\centering
\begin{subfigure}[t]{.49\textwidth}
    \centering
      \includegraphics[scale=.95]{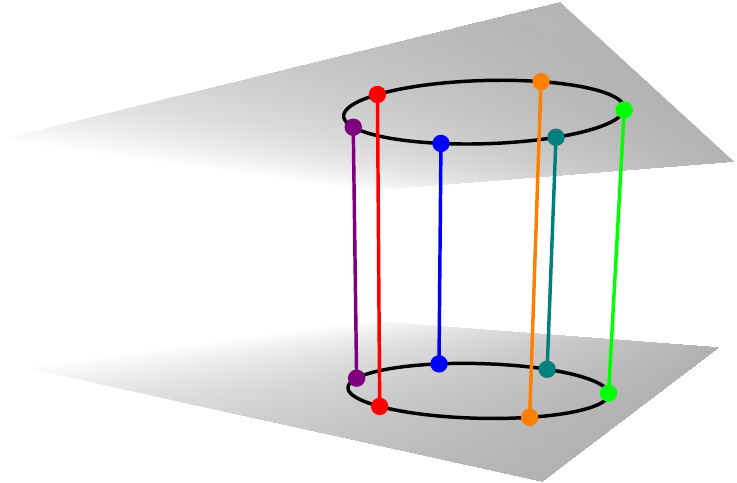}
      \caption{Case $n=0$: ${\gcd(n,N)}=6$ different branes.}
     \end{subfigure}
     \hfill
     \begin{subfigure}[t]{.49\textwidth}
      \centering
      \includegraphics[scale=.95]{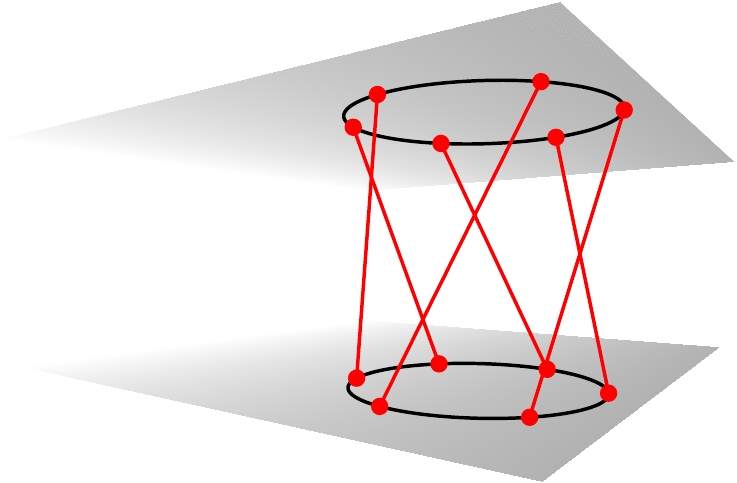}
      \caption{Case $n=1$: just ${\gcd(n,N)}=1$ brane.}
     \end{subfigure}
     \\[10mm]
    \begin{subfigure}[t]{.49\textwidth}
      \centering
      \includegraphics[scale=.95]{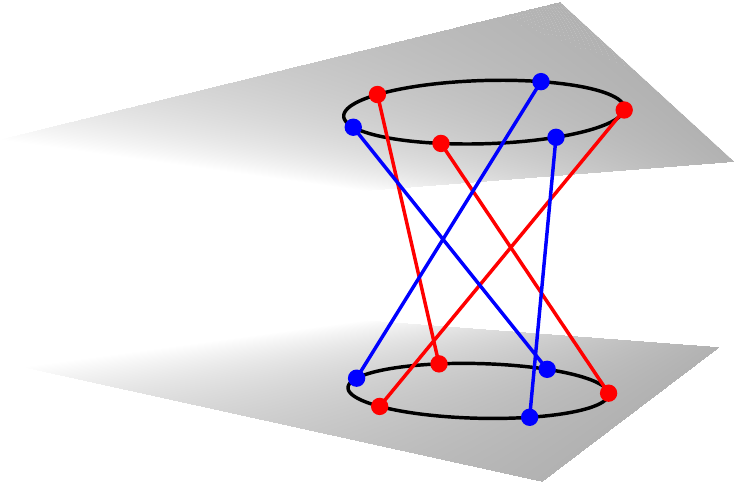}
      \caption{Case $n=2$: ${\gcd(n,N)}=2$ different branes.}
     \end{subfigure}
     \hfill
     \begin{subfigure}[t]{.49\textwidth}
      \centering
      \includegraphics[scale=.95]{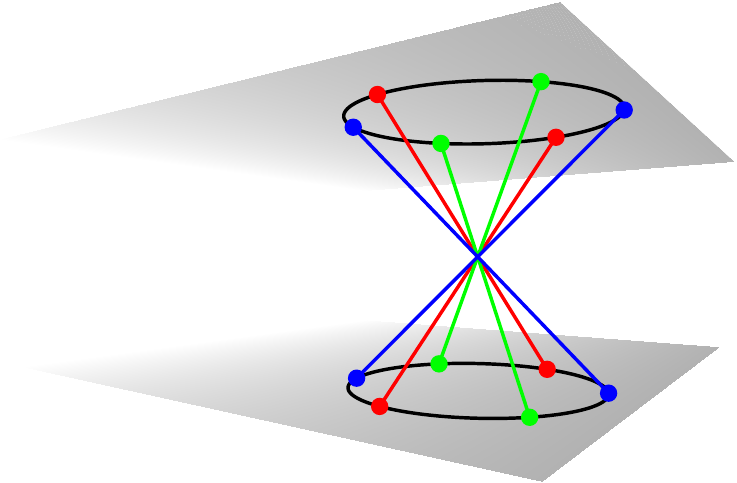}
      \caption{Case $n=3$: ${\gcd(n,N)}=3$ different branes.}
    \end{subfigure}
     \\[10mm]
    \begin{subfigure}[t]{.49\textwidth}
      \centering
      \includegraphics[scale=.95]{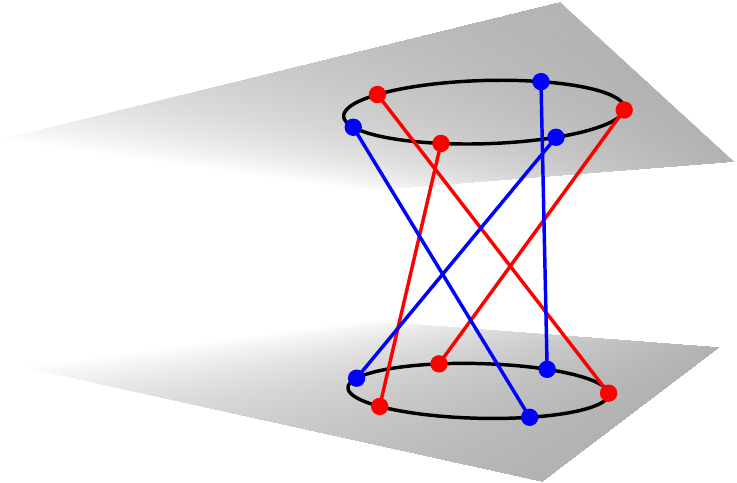}
      \caption{Case $n=4$: ${\gcd(n,N)}=2$ different branes.}
     \end{subfigure}
     \hfill
     \begin{subfigure}[t]{.49\textwidth}
      \centering
      \includegraphics[scale=.95]{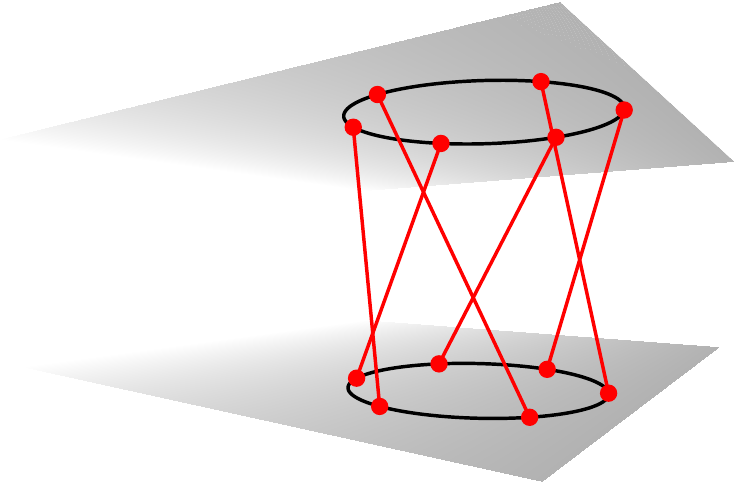}
      \caption{Case $n=5$: just ${\gcd(n,N)}=1$ brane.}
    \end{subfigure}
    \vspace*{1cm}
\caption{The case with $N=6$.
\label{fig:branes}}
\end{figure}

Let us note that, as far as the BPS state counting is concerned, it is enough to display the axio-dilaton profile. Indeed, the full supergravity solution will also include a non-trivial backreacted metric on $\bC$ \cite{Greene:1989ya}. Far from the D7-branes, this results in a deficit angle which restricts the maximal number of 7-branes in global models to be 24. On the other hand, to count BPS states we only need a local solution and in this case $N$ can be arbitrary (see \eg{} \cite{Karch:2002sh, Ouyang:2003df, Benini:2006hh, Benini:2007gx, Benini:2007kg} for examples in other contexts). Moreover, our construction is only sensitive to the holomorphic data of the background, here the axio-dilaton, and not to the metric which is a D-term deformation.

Next, we turn on the mass parameters $u_\alpha$ in a way compatible with the twisted geometry. For $\epsilon\neq0\mod1$, periodicity around the B-cycle of $T^2$ imposes constraints on $u_\alpha$. The simplest allowed choice is
\be
\label{configuration twisted masses}
u_\alpha = e^{2\pi i \alpha/N} u_{(0)} \qquad\qquad \text{for $\alpha = 1, \dots, N$}
\ee
and generic $u_{(0)}\in \bC$. This is a configuration where the branes homogeneously distribute on a circle around the origin. See Figure~\ref{fig:branes} for a pictorial representation of the various cases when $N=6$. From the field theory point of view, twisted masses are in general not compatible with the SUSY background that gives rise to the elliptic genus, because they are charged under the (left-moving) R-symmetry for which we turn on a background flat connection. However the special choice (\ref{configuration twisted masses}) is invariant under a combination of R-symmetry rotation and Weyl transformation within $\SU(N)$.

The elliptic genus does not depend on the twisted masses, therefore we can safely evaluate it for $u_\alpha$ as in (\ref{configuration twisted masses}). Because of the twist, the $N$ segments of D7-branes organise themselves into $\gcd(n,N)$ disconnected branes, each made of $N/\gcd(n,N)$ segments (see Figure~\ref{fig:branes}). Notice that these numbers are correct even in the case of no twist, $n=0$, in which the $N$ D7's are simply taken apart. The twisted geometry has a $\bZ_{N/\gcd(n,N)}$ symmetry, therefore if the number $k$ of D1-branes is not a multiple of that, they cannot be moved from the origin to the worldvolumes of the D7's. This reproduces the condition in (\ref{eq:ZkN}).

Finally, taking into account that each D7-brane is made of $N/\gcd(n,N)$ segments and so its worldvolume should be rescaled, we are left with a system of $\gcd(n,N)$ decoupled D7-branes, with a total of $k\gcd(n,N)/N$ D1-branes per segment to be distributed among the D7's. This is precisely the content of (\ref{eq:ZkN}), or its generating function (\ref{eq:Ztot}), up to the sign which is fixed by the R-symmetry anomaly. The extreme cases $n=0$ and $n=1$ are easier to understand.

\subsection{M-theory graviton index derivation: \\ An exercise on ``Membranes and Sheaves''}
\label{sec: M-theory}

We can give a geometric interpretation to the expression \eqref{eq:facttrig} in the realm of M-theory. This can be done as an exercise on \cite{Nekrasov:2014nea}.

Let us study our D-brane system from the viewpoint of M-theory. A bound state of $N$ D6-branes and $k$ D0-branes on $S^1$ can be lifted to an 11-dimensional bound state of $k$ gravitons on $S^1\times {\mathbb C}^3\times \mathrm{TN}_N$, where $\mathrm{TN}_N$ is
the $N$-center Taub-NUT space \cite{Sen:1997js,Hull:1997kt}. 
The $\Omega$-deformation of this lift is a twisted equivariant fibration, which has been considered in \cite{Nekrasov:2014nea}. Essentially, the toric space $\bC^3 \times \mathrm{TN}_N$ is rotated by an action of $U(1)^5$ as we circle around $S^1$, with a BPS constraint that the diagonal element does not act.

In the special case $N=1$ \cite{Nekrasov:2009jap}, the 11-dimensional lift contains a single-center Taub-NUT space whose topology is the same as $\bC^2$. Upon $\Omega$-deformation, the BPS graviton states localise towards the center of $\mathrm{TN}_1$ and become insensitive to the fact that its metric is different from that of $\bC^2$. Therefore, one can compute the BPS index of gravitons on the $\Omega$-deformed space by looking at the near-core geometry $\bC^3 \times \bC^2 \cong \bC^5$. The index of BPS single-particle graviton states (plus anti-BPS states) turns out to be \cite{Nekrasov:2009jap, Nekrasov:2014nea}
\be
F^{(11)}_1(q_1,q_2,q_3,q_4,q_5)=\frac{\sum_{i=1}^5 q_i}{\prod_{i=1}^5(1-q_i)}+
\frac{\sum_{i=1}^5 q_i^{-1}}{\prod_{i=1}^5(1-q_i^{-1})} \;.
\ee
For $\prod_{i=1}^5 q_i=1$, it can be decomposed as 
\be
F^{(11)}_1(q_1,q_2,q_3,q_4,q_5) = F^{(6)}(q_1,q_2,q_3) + F^{(6)}\big(q_1^{-1},q_2^{-1},q_3^{-1}\big) + \mc{F}_1(q_1,q_2,q_3;v) \;,
\ee
where
\bea
F^{(6)}(q_1,q_2,q_3) &= \frac{q}{\prod_{i=1}^3(1-q_i)} \;, \\
\mc{F}_1(q_1,q_2,q_3;v)&=\frac{\prod_{i=1}^3(1-q/q_i)}{\prod_{i=1}^3(1-q_i)} \times \frac{1}{(1-q^{1/2}v)(1-q^{1/2}v^{-1})} \;,
\eea
we set $q=q_1q_2q_3$ and solved $q_4=vq^{-1/2}$ and $q_5=v^{-1}q^{-1/2}$. One can interpret $F^{(6)}$ as the perturbative contribution to the free energy of the 7-dimensional theory on the D6-brane on $S^1 \times \bC^3$, and $\cF_1$ as the instanton part. In fact, $\cF_1$ is precisely the single-particle seed of the plethystic exponential in (\ref{eq:nekans}).

We can extend the computation of the BPS single-particle graviton index to the case $N>1$. As we said, the 11-dimensional lift of the D0/D6 system is a bound state of gravitons on $S^1 \times \bC^3 \times \mathrm{TN}_N$, and after $\Omega$-deformation this becomes a fibration of $\bC^3 \times \mathrm{TN}_N$ on $S^1$. Because the $\Omega$-deformation localises the graviton states around the origin of $\mathrm{TN}_N$, we can safely substitute $\mathrm{TN}_N$ by its near-core geometry, the orbifold space $\bC^2/\bZ_N$.

The index of BPS single-particle graviton states (plus anti-BPS states) on $\bC^3 \times [\bC^2/ \bZ_N]$ is easily obtained by projecting to the $\bZ_N$-invariant sector:
\be
F^{(11)}_N(q_1,q_2,q_3,q_4,q_5) = \frac{1}{N} \sum_{a=1}^N F^{(11)}_1\l(q_1,q_2,q_3,q_4^{(a)},q_5^{(a)}\r) \;,
\ee
where the fugacities along the orbifold directions are
\be
q_4^{(a)} = \omega^{(a)} v^{1/N} q^{-1/2} \;, \qquad\qquad q_5^{(a)} = \omega^{(-a)} v^{-1/N} q^{-1/2} \;,
\ee
and $\omega^{(a)}=e^{2\pi i a/N}$. To isolate the instanton counting factor, we subtract from the free energy the 7-dimensional perturbative contribution, and notice that $F^{(6)}$ is invariant under the $\bZ_N$ action. Setting
\be
F^{(11)}_N(q_1,q_2,q_3,q_4,q_5) = F^{(6)}(q_1,q_2,q_3) + F^{(6)}\left( q_1^{-1},q_2^{-1},q_3^{-1} \right) + {\mathcal F}_N(q_1,q_2,q_3;v) \;,
\ee
we obtain
\be
\label{enne-1}
\mc{F}_N(q_1,q_2,q_3;v) = \frac{\prod_{i=1}^3(1-q/q_i)}{\prod_{i=1}^3(1-q_i)} \times \frac{1}{N}\sum_{a=1}^N
\frac{1}{\big( 1- \omega^{(a)} q^{1/2} v^{1/N} \big) \big( 1 - \omega^{(-a)} q^{1/2} v^{-1/N} \big)} \;.
\ee
After resumming the last factor,%
\footnote{A convenient way to perform the sum is the following. Consider the function
$$
f(z) = \frac{1}{z^N-v}\cdot\frac{1}{z}\cdot\frac{1}{(1-q^{1/2}z)(1-q^{1/2}z^{-1})} \;,
$$
which has $N+2$ poles: at $z = v^{1/N}\omega^{(a)}$, $z=q^{1/2}$ and $z=q^{-1/2}$. Computing the residues and using that their sum is zero, one obtains the desired formula.}
we obtain
\be
\label{eq:resum}
\mc{F}_N(q_1,q_2,q_3;v) = \frac{\prod_{i=1}^3(1-q/q_i)}{\prod_{i=1}^3(1-q_i)}
\times \frac{q^N-1}{q-1} \times
\frac{1}{\big( 1-q^{N/2}v \big) \big( 1-q^{N/2}v^{-1} \big)} \;.
\ee
This is precisely the single-particle seed of the plethystic exponential in \eqref{eq:facttrig}.

\section{Free field representation of matrix integrals}
\label{sec: free field}

In this section we give a representation of the elliptic genus partition function in terms of chiral free bosons on the torus. The very existence of such a representation indicates that the elliptic vertex algebra, \ie{} the algebra of chiral vertex operators on the torus, might act on the cohomology of the moduli spaces that we have been studying so far and offer the language to detect a link to integrable systems in the spirit of the BPS/CFT correspondence \cite{Nekrasov:2015wsu}.

The rational case in dimension $0$ has a well-known free field representation in terms of chiral free bosons on the plane \cite{Kazakov:1998ji, Babelon:2003qtg}. In the following we will represent the grand canonical partition function for the elliptic genera as a combination of two factors: the torus (chiral) correlator of an exponentiated integrated vertex (whose power expansion reproduces the contributions from multiplets in the adjoint representation), and a linear source (that reproduces the contributions from multiplets in the fundamental representation).

It is well-known that an off-shell formulation of the chiral boson is difficult, therefore we will define it on-shell in the following way. Consider the usual free massless scalar boson two-point function
\be
\label{eq:t2_prop}
\big\langle \phi(u,\bar{u}) \, \phi(w,\bar{w}) \big\rangle_{T^2} = \log G(u,\bar{u};w,\bar{w})
\ee
where
\be
G(u,\bar{u};w,\bar{w}) = e^{- \f{2\pi}{\tau_2} \left( \rule{0pt}{.6em} \im( u-w ) \right)^2} \left\lvert\f{\jt{u-w}}{2\pi\eta(\tau)^3}\right\rvert^2 \;.
\ee
Here $\tau_2 = \im\tau$. Using this propagator, one computes the elliptic vertex algebra and the correlation functions of vertex fields of the usual type $\no{e^{\lambda\phi}}\,$.
A generic higher-point correlation function is the product of three factors: a holomorphic (in $u$ and $w$) contribution proportional to a product of functions $\theta_1$, an anti-holomorphic contribution proportional to $\bar\theta_1$'s, and a mixed contribution proportional to a product of exponentials. If the last term cancels out, then we can define---up to a pure $c$-number phase---the chiral projection of the correlation function by picking the holomorphic contribution.

Let us consider the following vertex operator:
\be
\mc{V}_{\vec{\epsilon}}(u)= \; \prod_{i=1}^7\no{e^{\lambda_i\phi_i(u_{+i})}} \, \no{e^{-\lambda_i\phi_i(u_{-i})}} \;,
\ee
where $\vec{\lambda}=(i,i,i,i,1,1,1)$ and
\bea
  u_{\pm i}&= u \pm \f{\tilde{\epsilon}_i}{2}\;, & \tilde{\epsilon}_1&= \epsilon_1\;, & \tilde{\epsilon}_2&=\epsilon_2\;, & \tilde{\epsilon}_3&=\epsilon_3\;, \\
  \tilde{\epsilon}_4&=\epsilon\;, & \tilde{\epsilon}_5&= \epsilon_{12}\;, & \tilde{\epsilon}_6 &= \epsilon_{13}\;, & \tilde{\epsilon}_7&=\epsilon_{23}\;,
\eea
are the vertices of two cubes with sides $\pm\epsilon_i/2$. At each vertex we placed one of $7$ non-interacting scalar fields on the torus with normalised two-point function
\be
\big\langle \phi_i(u,\bar{u}) \, \phi_j(w,\bar{w}) \big\rangle_{T^2}=\delta_{ij}\log G(u,\bar{u}; w, \bar{w}) \;.
\ee
Using Wick's theorem it is straightforward to find
\bea
  \mc{V}_{\vec\epsilon}(u) &= \prod_{i=1}^7 \Big[ G\big(u_{+i}, \bar u_{+i}; u_{-i}, \bar u_{-i}\big) \Big]^{\lambda_i^2} \, \no{\mc{V}_{\vec\epsilon}(u)}\\
  &= \l\lvert \f{2\pi\eta^3(\tau) \, \jt{\epsilon_{12}} \jt{\epsilon_{13}}\jt{\epsilon_{23}}}{\jt{\epsilon_1}\jt{\epsilon_2}\jt{\epsilon_3}\jt{\epsilon}}\r\rvert^2\no{\mc{V}_{\vec{\epsilon}}(u)}\;,\label{eq:Vnorm}
\eea
where, in the second line, the exponent of the imaginary parts squared cancels since
\be
\label{eq:remarkable}
\sum\nolimits_{i=1}^7\lambda_i^2 \, \big( \im(\tilde{\epsilon}_i) \big)^2 = 0 \;.
\ee
Again, using Wick's theorem, we find:
\begin{multline}
\no{e^{\lambda_i\phi_i(u_{+i})}e^{-\lambda_i\phi_i(u_{-i})}}\no{e^{\lambda_j\phi_j(u_{+j})}e^{-\lambda_j\phi_j(u_{-j})}} \, = {} \\
{} = \l[\f{G\l(u_{+i},\bar{u}_{+i};u_{-j},\bar{u}_{-j}\r)G\l(u_{-i},\bar{u}_{-i};u_{+j},\bar{u}_{+j}\r)}{G\l(u_{+i},\bar{u}_{+i};u_{+j},\bar{u}_{+j}\r)G\l(u_{-i},\bar{u}_{-i};u_{-j};\bar{u}_{-j}\r)}\r]^{\delta_{ij}\lambda_i\lambda_j}\times {} \\
{}  \times \no{e^{\lambda_i\phi_i(u_{+i})}e^{-\lambda_i\phi_i(u_{-i})}e^{\lambda_j\phi_j(u_{+j})}e^{-\lambda_j\phi_j(u_{-j})}} \;.
\end{multline}
The factor in square brackets, when, $i=j$ is
\begin{equation}
\l\lvert\f{\jt{u-v+\tilde{\epsilon}_i}\jt{u-v-\tilde{\epsilon}_i}}{\theta_1^2(\tau | u - v)}\r\rvert^{2\lambda_i^2}e^{-\f{4\pi}{\tau_2}\lambda_i^2\l(\im(\tilde{\epsilon}_i)\r)^2} \;,
\end{equation}
by which it follows that
\begin{multline}\label{eq:VV}
\langle\no{\mc{V}_{\vec{\epsilon}}(u)}\no{\mc{V}_{\vec{\epsilon}}(w)} \rangle = {} \\
{} = \l\lvert \f{ \theta_1^2(\tau |u-w) \, \jt{u-w-\epsilon_{12}} \jt{u-w-\epsilon_{13}} \jt{u-w-\epsilon_{23}}}{\jt{u-w+\epsilon_{1}} \jt{u-w+\epsilon_{2}}\jt{u-w+\epsilon_{3}}\jt{u-w-\epsilon}}\r. \times {} \\
{} \l. \times \f{\jt{u-w+\epsilon_{12}}\jt{u-w+\epsilon_{13}}\jt{u-w+\epsilon_{23}}}{\jt{u-w-\epsilon_{1}}\jt{u-w-\epsilon_{2}}\jt{u-w-\epsilon_{3}}\jt{u-w+\epsilon}} \r\rvert^2 .
\end{multline}
Notice that, again because of eq.~\eqref{eq:remarkable}, the exponent of the imaginary part squared cancels in \eqref{eq:VV} and we can define its holomorphic projection as
\begin{multline}\label{eq:VVholo}
\langle\no{\mc{V}_{\vec{\epsilon}}(u)}\no{\mc{V}_{\vec{\epsilon}}(w)}\rangle_{\mr{hol.}} = {} \\
{} = \f{\theta_1^2(\tau |u-w) \, \jt{u-w-\epsilon_{12}}\jt{u-w-\epsilon_{13}}\jt{u-w-\epsilon_{23}}}{\jt{u-w+\epsilon_{1}}\jt{u-w+\epsilon_{2}}\jt{u-w+\epsilon_{3}}\jt{u-w-\epsilon}} \times {} \\
{} \times \f{\jt{u-w+\epsilon_{12}}\jt{u-w+\epsilon_{13}}\jt{u-w+\epsilon_{23}}}{\jt{u-w-\epsilon_{1}}\jt{u-w-\epsilon_{2}}\jt{u-w-\epsilon_{3}}\jt{u-w+\epsilon}} \;,
\end{multline}
which is the contribution of single modes in the adjoint.

The other term that we need, in order to give a free-boson representation of our matrix model, is the following source operator:
\begin{equation}
\label{H}
H=\f{1}{2\pi i} \oint_\Gamma \partial\phi_4(w) \, \omega(w) \d w \;,
\end{equation}
where $\omega$ is a locally analytic function in the inner region bounded by the contour $\Gamma$. The contour $\Gamma$ is chosen to be a closed path around $w=0$ encircling all $u_{\pm i}$ for $i=1,\dots,7$ where $u=0$. Then we can compute%
\footnote{In the following formula we can trade $e^H$ with $\no{e^H}$ since $\omega$ is holomorphic inside $\Gamma$. Indeed, we have that $\no{e^H} \, = e^{\mf{N}} e^H$, where the normal ordering operator $\mf{N}$ is defined as
\begin{equation}
\mf{N} = \int\d^2z \, \d^2w\, \bigl\langle \phi(z,\bar{z}) \, \phi(w,\bar{w}) \bigr\rangle \, \f{\delta}{\delta\phi(z,\bar{z})} \, \f{\delta}{\delta\phi(w,\bar{w})} \;.
\end{equation}
We consider now
\begin{equation}
\mf{N} \, e^H = \frac1{(2\pi i)^2} \oint_\Gamma\d u \, \omega(u) \oint_\Gamma\d u' \omega(u') \, \partial_u\partial_{u'} \bigl\langle \phi(u,\bar{u})\phi(u',\bar{u}') \bigr\rangle \, e^H
= - \oint_\Gamma \d u \, \omega(u) \, \partial\omega(u) \, e^H = 0 \;.
\end{equation}
This implies our claim.
}
\be
e^H\no{e^{\lambda_j\phi_j(u_{+j})}e^{-\lambda_j\phi_j(u_{-j})}} = e^{W}\no{e^He^{\lambda_j\phi_j(u_{+j})}e^{-\lambda_j\phi_j(u_{-j})}} \;,
\ee
where
\begin{align} \label{eq:HVpartial}
W &= \delta_{4j} \lambda_j \f{1}{2\pi i}\oint_\Gamma\d w \, \omega(w) \Bigl[ \partial_w \bigl\langle \phi_4(w) \, \phi_j(u_{+j}) \bigr\rangle - \partial_w \bigl\langle \phi_4(w) \, \phi_j(u_{-j}) \bigr\rangle \Bigr] \\
  &=\delta_{4j} \lambda_j \l[\f{1}{2\pi i}\oint_\Gamma\d w \, \omega(w) \Bigl[ \zeta_{\mr{W}}(w-u_{+j})-\zeta_{\mr{W}}(w-u_{-j}) \Bigr] - \f{2i}{\tau_2}\oint_\Gamma\d w \, \omega(w)\im(\tilde{\epsilon}_j)  \r]\;, \nn
\end{align}
where we introduced the Weierstrass $\zeta$ function $\zeta_{\mr{W}}(u)=\partial\log\jt{u}$ which has a simple pole around the origin:
\begin{equation}
  \zeta_{\mr{W}}(u)= \f{1}{u} + \text{holomorphic in $u$} \;.
\end{equation}
The second term in the last line of \eqref{eq:HVpartial} is zero since $\omega$ is holomorphic inside $\Gamma$. It follows that
\begin{equation}\label{eq:HVpre}
\bigl\langle e^{H} \no{\mc{V}_{\vec{\epsilon}}(u)} \bigr\rangle= e^{\f{1}{2\pi}\oint_{\Gamma}\l[(w-u_{+4})^{-1}-(w-u_{-4})^{-1} \r]\omega(w)\d w}
  = e^{i\omega(u+\epsilon/2)-i\omega(u-\epsilon/2)} \;.
\end{equation}
Choosing (up to an irrelevant additive constant)
\begin{equation}
  \omega(u)=i\sum_{\alpha=1}^{N}\log \jt{u+z_\alpha-\f{\epsilon}{2}}\, ,
\end{equation}
which is holomorphic inside $\Gamma$ for generic values%
\footnote{The branch cuts of the logarithms generically extend outside the contour.}
of the Cartan parameters $\{z_\alpha\}$,  eq.~\eqref{eq:HVpre} reads
\begin{equation}\label{eq:HV}
\bigl\langle e^{H}\no{\mc{V}_{\vec{\epsilon}}(u)} \bigr\rangle_{\mr{hol.}} = \prod_{\alpha=1}^N\f{\jt{u+z_\alpha-\epsilon}}{\jt{u+z_\alpha}} \;.
 \end{equation}
Moreover notice that, since only the chiral part of the scalar boson enters eq.~\eqref{H}, eq.~\eqref{eq:HV} is already holomorphic, so we add the subscript ``hol.'' without further ado. Now using \eqref{eq:VVholo} and \eqref{eq:HV}, we can expand
\begin{multline}\label{eq:mmcc}
\bigl\langle e^{H} e^{{v}\oint_{\mc{C}}{\mc{V}_{\vec{\epsilon}}(u)} \, \d u} \bigr\rangle_{\mr{hol.}}
=
\sum_{k=0}^\infty\f{{v}^k}{k!}\l[ \f{2\pi\eta^3(\tau) \, \jt{\epsilon_{12}}\jt{\epsilon_{13}}\jt{\epsilon_{23}}}{\jt{\epsilon_1}\jt{\epsilon_2}\jt{\epsilon_3}\jt{\epsilon}} \r]^k \times {} \\
{} \times \oint_{\mc{C}}\d u_1 \dots\oint_{\mc{C}}\d u_k
     \prod_{i=1}^k \prod_{\alpha=1}^N
     \f{\jt{u_i+z_\alpha-\epsilon}}{\jt{u_i+z_\alpha}} \times {} \\
{} \times\prod_{\substack{i,j=1\\i\neq j}}^k
                             \f{\jt{u_{ij}}\jt{u_{ij}-\epsilon_{12}}\jt{u_{ij}-\epsilon_{13}}\jt{u_{ij}-\epsilon_{23}}}{\jt{u_{ij}+\epsilon_1}\jt{u_{ij}+\epsilon_2}\jt{u_{ij}+\epsilon_3}\jt{u_{ij}-\epsilon}} \;.
\end{multline}
Notice that the prefactor in the first line arises from the fact that in the l.h.s. $\mc{V}_{\vec\epsilon}$ is present \emph{without} normal ordering---see the holomorphic part of \eqref{eq:Vnorm}. Comparing eqs.~\eqref{eq:grcan_ellgen} and \eqref{eq:mmcc} we realise that
 \begin{equation}
   Z^{(N)}(v) = \bigl\langle e^{H}e^{{v}\oint_{\mc{C}}\mc{V}_{\vec{\epsilon}}(u) \, \d u} \bigr\rangle_{\mr{hol.}} \;,
 \end{equation}
provided the contour $\mc{C}$ is the one specified by the JK prescription. We remark that the function defined through $H$ can be lifted to $T^2$ in cases in which the $\mr{R}$-symmetry is not anomalous, that is $\epsilon\in\mb{Z}$.

\section{Conclusion and outlook}
\label{sec: conclusions}

In this paper we have studied the dynamics of the D1/D7 brane system on an elliptic curve $T^2$. The effective dynamics of the D1-branes is a gauged linear sigma model, whose elliptic genus computes the equivariant elliptic genus of rank-$N$ sheaves on $\mathbb{C}^3$. We computed the elliptic genus using the supersymmetric localisation formula of \cite{Benini:2013nda,Benini:2013xpa}, which reduces the problem to a Jeffrey-Kirwan residue \cite{JK95} evaluation. We showed that the poles contributing to the integral are in one-to-one correspondence with $N$-coloured plane partitions. The proof requires to disentangle some subtleties related to the desingularisation of the integrand, that to the best of our knowledge were not previously discussed in the literature. Details on this are reported in Appendix~\ref{app: technical details}. One important feature of the two-dimensional sigma model is that it is gapped in the IR and, due to anomalies, only has a discrete axial R-symmetry. From the mathematical viewpoint this means that the complex (equivariant) parameter needs to take special discrete values. The elliptic genus takes a particularly simple form given by (\ref{eq:ZkN}), that can be interpreted in terms of D1/D7-brane bound-state counting in the strongly coupled IIB superstring/F-theory context, as discussed in Section  \ref{sec: F-theory}.

We also thoroughly studied dimensional reductions of the sigma model to $\mc{N}=4$ gauged quantum mechanics (QM) and to a matrix model. The quantum-mechanical system is expected to compute K-theoretic rank-$N$ Donaldson-Thomas invariants. We analysed a conjectural plethystic exponential form for the QM partition function in (\ref{eq:facttrig}), which generalises the one conjectured in \cite{Nekrasov:2009jap} and proved in \cite{Nekrasov:2014nea}. The formula has a nice interpretation as the 11-dimensional supergravity (or M-theory) index on the background $S^1\times \mathbb{C}^3\times \mathbb{C}^2/\mathbb{Z}_N$ with $\Omega$-deformation, in agreement with the results of \cite{Nekrasov:2014nea}. Therefore, (\ref{eq:facttrig}) is a conjectural plethystic exponential formula for higher-rank equivariant Donaldson-Thomas invariants on $\mathbb{C}^3$. We underline that in the QM case the higher-rank result does not factorise in Abelian contributions, due to the presence of non-trivial twisted sectors under the orbifold. We instead confirm that the factorisation holds in the matrix model limit, as conjectured in \cite{Nekrasov:2009jap} and verified in \cite{Cirafici:2008sn, Szabo:2015wua}. The relevant formula for the matrix model case is (\ref{eq:factrat}), that we checked with our techniques up to 8\textsuperscript{th} order in the instanton expansion.

Finally, we studied a free field representation of the elliptic genus in terms of integrated vertex operators of chiral fields on the torus, whose chiral correlators reproduce the contribution of adjoint fields in the D1 gauge theory, and a source term, which is necessary to reproduce the fundamental multiplet contribution. This result generalise to the D1/D7 system the construction of \cite{Kazakov:1998ji} and point to the existence of an elliptic vertex algebra acting on the associated moduli space of sheaves, see \cite{Zenkevich:2017tnb} for recent progress in this direction. We also expect this result to prompt a constructive connection with integrable hierarchies, which would be very interesting to investigate.

Another natural direction for future work is the study of the D1/D7 system on more general toric geometries, such as the conifold, where a wall crossing phenomenon among different geometric phases of the moduli space is expected to arise, see \cite{Cirafici:2012qc} for a review. On such geometries, bound states including D2-branes become important, and a description of D2/D6 systems in terms of 3d Chern-Simons-matter theories \cite{Aganagic:2009zk, Benini:2009qs, Benini:2011cma} might turn useful. In our approach, the different phases should be related to different choices of the integration contour. Moreover, it would be interesting to investigate whether the factorisation property of the matrix model limit is spoiled on more general geometries.   

It would be also interesting to investigate along these lines the supersymmetric partition function on compact toric three-folds, as for example $\mathbb{P}^3$ or $\mathbb{P}^1\times \mathbb{P}^2$,
in order to compute topological invariants of higher-rank stable sheaves on them. Analogous computations in two complex dimensions have been performed in \cite{Bawane:2014uka, Bershtein:2015xfa, Bershtein:2016mxz}, while 
some results for three-folds already appeared in the mathematical literature \cite{Gholampour:2013hga}.

\section*{Acknowledgements} 

We would like to thank Michele Cirafici, Omar Foda and Kimyeong Lee for useful discussions. G.B., M.P. and A.T. would like the GGI INFN Institute for having provided a stimulating atmosphere where part of this work has been done.

The work of F.B. is supported in part by the MIUR-SIR grant RBSI1471GJ ``Quantum Field Theories at Strong Coupling: Exact Computations and Applications''.
The research of F.B., G.B. and of M.P. is supported in part by INFN via the Iniziativa Specifica ST\&FI.
The work of G.B. and M.P. is supported by the PRIN project ``Non-perturbative Aspects Of Gauge Theories And Strings''.
The research of A.T. is supported by INFN via the Iniziativa Specifica GAST and PRIN project ``Geometria delle variet\`a algebriche''.

\appendix

\section{Special functions}
\label{app: special functions}

First of all we define the modular parameter to be $p=e^{2\pi i \tau}$, with $\im\tau>0$. The $q$-Pochhammer symbol is defined as
\be
(y;p)_\infty = \prod_{k=0}^\infty(1-yp^k) \;.
\ee
The Dedekind eta function and a suitable theta function can be written as
\be
\eta(p) = p^{\f{1}{24}} (p;p)_\infty \;,\qquad\qquad  \theta(\tau | z) = (y;p)_\infty (py^{-1};p)_\infty \;,
\ee
where we set for convenience $y=e^{2\pi i z}$. The most ubiquitous function in this paper is the Jacobi theta function of the first kind:
\be
\theta_1(\tau | z) = i p^{\f{1}{8}}y^{-\f{1}{2}}(p;p)_\infty \theta(\tau | z)
\ee
such that $\theta_1(\tau|z) = - \theta_1(\tau|-z)$. Under shifts $z \mapsto z+a+b\tau$ with $a,b\in\mb{Z}$ of the argument, the function transforms as
\be
\label{eq:thetatrasf}
\theta_1 \big( \tau | z + a + b\tau \big) = (-1)^{a+b} \, e^{-2\pi i b z} \, e^{-i\pi b^2\tau} \, \theta_1(\tau | z) \;.
\ee
The function $\theta_1(\tau | z)$ has no poles, while simple zeroes occur for $z \in \mb{Z}+\tau\mb{Z}$. The residues of its inverse are
\be
\label{eq:theta_res}
\f{1}{2\pi i}\oint_{z=a+b\tau}\f{\d z}{\theta_1(\tau | z)} = \f{(-1)^{a+b} \, e^{i\pi b^2 \tau}}{2\pi \eta^3(\tau)} \;.
\ee
For small values of $p$ and $z$ we have
\be
\label{eq:theta_small}
\theta_1(\tau | z) \;\xrightarrow{p\rightarrow 0}\; 2p^{\f{1}{8}}\sin(\pi z) \;\xrightarrow{z\rightarrow 0}\; 2\pi p^{\f{1}{8}}z \;.
\ee

\section{Plethystic exponential}
\label{app: plethystic exp}

Let us define the plethystic exponential, following \cite{Feng:2007ur, Rastelli:2016tbz}. Given a function $f(x_1, \dts, x_n)$ of $n$ variables, such that it vanishes at the origin, $f(0, \dts, 0) = 0$, we set
\be
\label{eq:pletdef}
\pe{x_1,\ldots,x_n}{\rule{0pt}{1em} f(x_1,\dts,x_n)} \equiv \exp\l\{\sum_{r=1}^\infty \f{f(x_1^r,\dts,x_n^r)}{r}\r\} \;.
\ee
If $f$ is $C^\omega$ with expansion
\be
f(x_1,\dts,x_n) = \sum_{m_1,\dots,m_n=1}^\infty f_{m_1,\ldots,m_n} \, x_1^{m_1} \cdots x_n^{m_n} \;,
\ee
then \eqref{eq:pletdef} can be rewritten as
\be
\pe{x_1,\ldots,x_n}{\rule{0pt}{1em} f(x_1,\dts,x_n)} = \prod_{m_1,\ldots,m_n}^\infty \big( 1-x_1^{m_1} \cdots x_n^{m_n} \big)^{-f_{m_1,\ldots,m_n}} \;.
\ee

\section{Plane partitions}
\label{App:pp}

A list of integers $\pi^{(1)}=\{a_1,\dts,a_\ell\}$ such that $a_i\geq a_{i+1}$ and whose sum is a given integer $k$, is called a \emph{partition} of $k$. We define $|\pi^{(1)}| = k$. Partitions of $k$ are in one-to-one correspondence with Young diagrams with $k$ boxes. We call $\phi_k$ the number of partitions of $k$, and their generating function is
\be
\phi(v) \,\equiv\, \sum_{k=0}^\infty \phi_k\, v^k = \prod_{k=1}^\infty \frac1{1-v^k} = \pe{v}{ \frac v{1-v} } \;.
\ee

We can introduce a partial order relation $\succeq$ among partitions: we say that $\pi^{(1)}_1\succeq\pi^{(1)}_2$ if the Young diagram representing $\pi^{(1)}_1$ ``covers'' the one representing $\pi^{(1)}_2$. We can then iterate the process. We define a \emph{plane partition} of $k$ as a collection of Young diagrams
\be
\pi^{(2)} = \l\{\pi^{(1)}_1,\ldots,\pi^{(1)}_\ell \r\} \qquad\text{such that } \pi^{(1)}_i\succeq \pi^{(1)}_{i+1} \qquad \text{and } |\pi^{(2)}| \equiv \sum_{r=1}^\ell |\pi^{(1)}_r| = k \;.
\ee
We can imagine $\pi^{(2)}$ as a pile of $\ell$ Young diagrams placed one on top of the other. We call $\Phi_k$ the number of plane partitions of $k$. Their generating function $\Phi$ was found by MacMahon to be
\be
\Phi(v) \,\equiv\, \sum_{k=0}^\infty \Phi_k \, v^k = \prod_{k=1}^\infty \frac1{(1-v^k)^k} = \pe{v}{ \frac v{(1-v)^2} } \;.
\ee
In this paper we denote a plane partition simply by $\pi$ without any superscript.

A \emph{coloured plane partition} is a collection of $N$ plane partitions. The generating function of the numbers $\Phi^{(N)}_k$ of coloured plane partitions of $k$ is simply the $N$-th power of the generating function of uncoloured plane partitions:
\be
\sum_{k=0}^\infty \Phi^{(N)}_k\, v^k = \Phi(v)^N \;.
\ee
For instance:
\be
\Phi_0^{(N)} = 1 \;,
\quad
\Phi_1^{(1)}=N \;,
\quad
\Phi_2^{(N)}=3N + \binom{N}{2} \;,
\quad
\Phi_3^{(N)}=6N+6\binom{N}{2}+\binom{N}{3} \;.
\ee

 \begin{figure}
  \centering
  \begin{subfigure}[t]{.32\textwidth}
   \centering
   \includegraphics{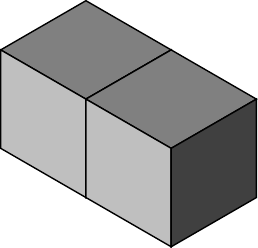}
   \caption{}
  \end{subfigure}
  \begin{subfigure}[t]{.32\textwidth}
   \centering
   \includegraphics{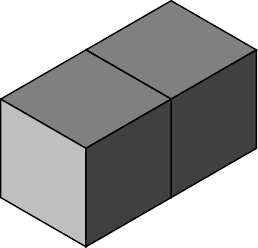}
   \caption{}
  \end{subfigure}
    \begin{subfigure}[t]{.32\textwidth}
   \centering
   \includegraphics{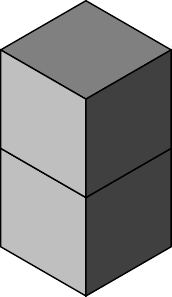}
   \caption{}
  \end{subfigure}
  \caption{Plane partitions of the case $k=2$.}\label{fig:2pp}
 \end{figure}

\section{Technical details}
\label{app: technical details}

\subsection{Canonical form of the charge matrix}
\label{app:canonical}

In order to have isolated solutions of \eqref{eq:ls}, $\ms{Q}$ must have non-vanishing determinant. This is possible if $f$ -- that is the number of $\vec{h}_j$'s which represent hyperplanes of type $H_F$ -- is greater or equal than one. In order to find a canonical form of $\ms{Q}$ we will use two moves:
\begin{itemize}
\item swap columns: this is equivalent to relabelling the $\beta$'s;
\item swap rows: this is equivalent to a Weyl transformation, i.e. to a permutation of $u$'s.
\end{itemize}
The algorithm to reach the canonical form goes as follows:
\begin{description}
\item[Step 1:] Choose $\vec{v}_1$, a vector of type $\vec{h}_F$ among $\vec{h}^{\ms{T}}_i$ with $i=1,\dots,k$. Shuffle rows so that the only non-vanishing entry of $\vec{v}_1$ sits at the first row. Shuffle the columns so that $\vec{v}_1$ is $\vec{h}_1^{\ms{T}}$.
\item[Step 2:] Choose $\vec{v}_2$ among $\vec{h}_i^{\ms{T}}$ with $i=2,\dots,k$ such that its first entry is non-vanishing. If there is no such a vector, go to {\bf Intermezzo}. The vector $\vec{v}$ will have another non-zero entry to maintain $\det\ms{Q}\neq 0$: shuffle the rows after the first so that the first two entries of $\vec{v}_2$ are non-zero while the other vanish. Shuffle the columns after the first so that $\vec{v}_2$ is $\vec{h}_2^{\ms{T}}$.
\item[Step $p$:] Choose $\vec{v}_p$ among $\vec{h}_i^{\ms{T}}$ with $i=p,\dots,k$ such that its first $p$ entry are not all vanishing. If there is no such a vector go to  {\bf Intermezzo}. The vector $\vec{v}_p$ will have another non-vanishing component after the $(p-1)$\textsuperscript{th} entry, otherwise $\vec{h}_1,\dots,\vec{h}_p$ would be linear dependent and $\det\ms{Q}=0$. Shuffle the rows after the $(p-1)$\textsuperscript{th} so that this non-vanishing value sits in the $p$\textsuperscript{th} entry. Shuffle the columns after the $(p-1)$\textsuperscript{th} so that $\vec{v}_p$ is $\vec{h}_p^{\ms{T}}$.
\item[Intermezzo:] After having chosen $k_1$ vectors $\vec{v}_1,\dots,\vec{v}_k$ (since they are in finite number) we are in the situation in which there are no more vectors $\vec{h}_i^{\ms{T}}$ with $i=k_1+1,\dots,k$ having the first $k_1$ entries not all vanishing. At this step the charge matrix looks like
\begin{equation}
  \ms{Q}= 
\begin{tikzpicture}[
  baseline,
  label distance=10pt 
]

\matrix [matrix of math nodes,left delimiter=(,right delimiter=),row sep=0.1cm,column sep=0.1cm] (m) {
      1 & \tilde{*}    &   \tilde{*} & \dots & \tilde{*}   & 0 & \dots & 0 \\
      0 & \pm 1 &   \tilde{*}  & \dots & \tilde{*}   & 0 & \dots & 0 \\
      0 & 0     &\pm 1 & \dots & \tilde{*}   & 0 & \dots & 0 \\
      \vdots & \vdots & \vdots & \ddots & \vdots & \vdots & \ddots & \vdots\\
        0 &  0    & 0    &   \dots   &\pm 1& 0 & \dots & 0 \\
        0 &  0    & 0    &  \dots   & 0  & \tilde{*} & \dots & \tilde{*} \\
      \vdots & \vdots & \vdots & \ddots & \vdots & \vdots & \ddots & \vdots\\
      0 &  0    & 0    &   0   & 0   & \tilde{*} & \dots & \tilde{*}\\ };

\draw[dashed] ($0.5*(m-1-5.north east)+0.5*(m-1-6.north west)$) -- ($0.5*(m-8-6.south east)+0.5*(m-8-5.south west)$);

\draw[dashed] ($0.5*(m-5-1.south west)+0.5*(m-6-1.north west)$) -- ($0.5*(m-5-8.south east)+0.5*(m-6-8.north east)$);

\node[
  fit=(m-1-1)(m-1-5),
  inner xsep=0,
  above delimiter=\{,
  label=above:$k_1$
] {};

\node[
  fit=(m-1-6)(m-1-8),
  inner xsep=0,
  above delimiter=\{,
  label=above:$k-k_1$
 ] {};

\node[
  fit=(m-1-8)(m-5-8),
  inner xsep=15pt,inner ysep=0,
  right delimiter=\},
  label=right:$k_1$
] {};

\node[
  fit=(m-6-8)(m-8-8),
  inner xsep=15pt,inner ysep=0,
  right delimiter=\},
  label=right:$k-k_1$
] {};

\end{tikzpicture}\;.
\end{equation}
Every $\tilde{*}$ represent a value that can be either $0$ or $\pm 1$ so that every column is a charge vector like \eqref{eq:charge_vector}.
\item [Steps from $k_1+1$ to $k_2$:] Repeat {\bf Steps} above on the right-bottom block with $f-1$ vectors $\vec{h}_i^{\ms{T}}$ representing hyperplanes of type $H_F$.
\item [Steps from $k_2+1$ to $k_f$:] Repeat {\bf Steps} above until there are no more vectors in the right-bottom block:
  \begin{equation}
    \sum_{q=1}^f k_q = k\;.
  \end{equation}
\item [Coda:] At the end of this procedure the charge matrix is block diagonal
  \begin{equation}\label{eq:Qblock}
    \ms{Q} = \diag(\ms{Q}_1,\dots,\ms{Q}_f)\;,
    \qquad
    \text{with}\;\,
     \ms{Q}_q=
     \begin{pmatrix}
       1    &   \tilde{*}   &   \tilde{*}   &  \dots  &    \tilde{*}   \\
       0    & \pm 1 &   \tilde{*}   &  \dots  &    \tilde{*}   \\
       0    &   0   & \pm 1 &  \dots  &    \tilde{*}   \\
       \vdots & \vdots & \vdots & \ddots & \vdots \\
       0    &   0   &  0    &  \dots  &    \pm 1
     \end{pmatrix}\;,
     \quad
     q=1,\dots, f\;.
  \end{equation}
\end{description}
Until here we did not use the condition $\beta_j>0$ as in \eqref{JK condition with beta's}. Since we have proven that $\ms{Q}$ is block diagonal we can impose block by block the condition of positivity of $\beta$'s:
\begin{equation}\label{eq:Qblock_problem}
  \ms{Q}_q\vec{\beta}_q = \vec{\eta}_q\;,
  \qquad
  q=1,\dots,f\;,
  \qquad\qquad
  \text{with }
  \vec{\beta}_q=
  \begin{pmatrix}
    \beta_{q,1}\\
    \vdots\\
    \beta_{q,k_q}
  \end{pmatrix}\;,
\end{equation}
where $\vec{\beta}_q$ is the part of $\vec{\beta}$ corresponding to the $q$\textsuperscript{th} block. The same is for $\vec{\eta}_q$. Comparing eq.~\eqref{eq:Qblock_problem} with eq.~\eqref{eq:Qblock} we see that the solution for positive $\beta_{q,k_q}$ is
\begin{equation}
  \beta_{q,k_q}=1\;,
  \qquad\qquad\qquad
   \ms{Q}_q=
     \begin{pmatrix}
       1    &   \tilde{*}   &   \tilde{*}   &  \dots  &    {*}   \\
       0    & \pm 1 &   \tilde{*}   &  \dots  &    {*}   \\
       0    &   0   & \pm 1 &  \dots  &    {*}   \\
       \vdots & \vdots & \vdots & \ddots & \vdots \\
       0    &   0   &  0    &  \dots  &    + 1
     \end{pmatrix}\;,
\end{equation}
that is, we have restricted the values of the last columns of $\ms{Q}$: the values of $*$ can be just either $0$ or $-1$. We can go ahead with this procedure: in order to do so we introduce the following notation: $\ms{Q}_q^{(i)}$ indicates the matrix $\ms{Q}_q$ with the last $i$ rows and $i$ columns removed; while $\vec{v}^{(i)}$ denotes the vector $\vec{v}$ with the last $i$ entries removed. From eq.~\eqref{eq:Qblock_problem} follows
\begin{equation}
  \ms{Q}_q^{(1)}\vec{\beta}_q^{(1)} = \vec{\eta}_q^{(1)} - \beta_{q,k_q}\vec{\ms{q}}_{q,k_q}^{(1)}\;,
\end{equation}
where we introduced $\vec{\ms{q}}_{q,i}$ as the $i$\textsuperscript{th} column vector of $\ms{Q}_q$. We see that on the r.h.s. we have a vector which is made of all $1$ except an entry, which is $2$. From this fact,  we can infer as above that
\begin{equation}
  \beta_{q,k_q-1} \geq 1\;,
   \qquad\qquad\qquad
   \ms{Q}_q^{(i)}=
     \begin{pmatrix}
       1    &   \tilde{*}   &   \tilde{*}   &  \dots  &    {*}   \\
       0    & \pm 1 &   \tilde{*}   &  \dots  &    {*}   \\
       0    &   0   & \pm 1 &  \dots  &    {*}   \\
       \vdots & \vdots & \vdots & \ddots & \vdots \\
       0    &   0   &  0    &  \dots  &    + 1
     \end{pmatrix}\;.
\end{equation}
The argument above can be easily iterated:
\begin{equation}
  \ms{Q}_q^{(i)}\vec{\beta}_q^{(i)} = \vec{\eta}_q^{(i)} - \sum_{j=0}^{i-1}\beta_{q,k_q-j}\vec{\ms{q}}_{q,k_q-j}^{(i)}\;,
\end{equation}
at every step we discover that $\beta_{q,k_q-j} \geq \beta_{q,k_q-j+1}$. Therefore we have that
\begin{equation}\label{eq:Qfinal}
   \ms{Q} = \diag(\ms{Q}_1,\dots,\ms{Q}_f)\;,
     \qquad
     \text{with}\;\,
     \ms{Q}_q=
     \begin{pmatrix}
       1    & -1  &  {*}   &  \dots  &    {*}   \\
       0    & +1 &   {*}   &  \dots  &    {*}   \\
       0    &   0   &  +1 &  \dots  &     {*}   \\
       \vdots & \vdots & \vdots & \ddots & \vdots \\
       0    &   0   &  0    &  \dots  &   +1
     \end{pmatrix}\;,
     \quad
     q=1,\dots, f\;,
   \end{equation}
   and
   \begin{equation}
      k_q=\beta_{q,k_q} \leq \beta_{q,k_q-1} \leq \dots \leq \beta_{q,2} \leq \beta_{q,1} =1\;.
   \end{equation}
   The fact that $\beta_{q,k_q}=k_q$ can be argued summing all the rows in eq.~\eqref{eq:Qblock_problem} and plugging the result \eqref{eq:Qfinal}.
   
With this new information, we can write eq.~\eqref{eq:ls} block by block
\begin{equation}
     \ms{Q}_q^{\ms{T}}\vec{u}_q = \vec{d}_q \;, \qquad\quad
     \text{with\footnotemark} \qquad
     \vec{u}_q =
     \begin{pmatrix}
       u_{q,1}\\
       \vdots\\
       u_{q,k_q}
     \end{pmatrix}
     \quad\text{ and }\quad
     \vec{d}_q=
     \begin{pmatrix}
       d_{q,1}\\
       \vdots\\
       d_{q,k_q}
     \end{pmatrix}\;.
\end{equation}
\footnotetext{We are relabelling the components of $\vec{u}$ and $\vec{d}$: $u_{q,i} = u_{i+\sum_{r=1}^{q-1}k_r}$ and $d_{q,i} = d_{i+\sum_{r=1}^{q-1}k_r}$.}
An important consequence of the form of $\ms{Q}$ in eq.~\eqref{eq:Qfinal} is that
\begin{equation}
     u_{q,j} - u_{q,i} \in -\epsilon_1\mb{Z}_+ -\epsilon_2\mb{Z}_+ -\epsilon_3\mb{Z}_+\;,
     \qquad\quad
     \text{for $j > i$}\;.
\end{equation}

\subsection{Desingularisation procedure}
\label{app:desing}

Let $I$ be the integrand in eq.~\eqref{elliptic genus N=1}. Suppose that the JK prescription implies to take the residue for $\{u_i\rightarrow\hat{u}_i\}_{i=1}^k$. It is always possible to order the factors of $I$ in the following way:%
\footnote{Since $\jt{\bullet}$ is odd, possible minus signs inside the argument can be reabsorbed in the $f_i$'s.}
\begin{equation}\label{eq:to_desing}
     I(\vec{u}) = \prod_{i=1}^kI_i(u_1,\dots,u_i)\;,
     \qquad
     I_i = \f{\prod_{c_i=1}^{C_i}\jt{u_i - u_{\gamma_{i,c_i}} + s_{i,c_i}}}{\prod_{a_i=1}^{A_i}\jt{u_i - u_{\alpha_{i,a_i}} + r_{i,c_i}}}f_i(u_1,\dots,u_i)\;,
\end{equation}
where $f$ contains all the factors which are both regular and non-zero for $\{u_i\rightarrow\hat{u}_i\}_{i=1}^k$, while in the fraction we put all the other ones. Thus, for $\{u_i\rightarrow\hat{u}_i\}_{i=1}^k$ there will be $A\equiv\sum_{i=1}^kA_i$ singular hyperplanes and $C\equiv\sum_{i=1}^kC_i$ zero hyperplanes. The interesting case is when $A \geq k$, since in the other cases, the residue is trivially vanishing. Then $\alpha_\bullet$ and $\gamma_\bullet$ are sequences such that $0 \leq \alpha_{i,a_i} \leq i$ and $0\leq\gamma_{i,c_i}\leq i$. In this way every $I_i$ depends only on $u_j$ with $j \leq i$. We allowed also to have $u_0\equiv 0$ in order to subsume all possible factors of Tab.~\ref{tab: sing} in the same form. Coefficients $r_{i,a_i}$ and $s_{i,c_i}$ are combination of $\epsilon$'s as in Tab.~\ref{tab: sing}. If $A = k$ we are in the \emph{regular} case of JK procedure and we can compute recursively 
   \begin{equation}
     \lim_{\{u_i\rightarrow\hat{u}_i\}_{i=1}^k}\f{I(\vec{u})}{\bigl( 2\pi\eta(\tau)^3 \bigr)^k\prod_{i=1}^k(u_i-\hat{u}_i)} = \bigl( 2\pi\eta(\tau)^3 \bigr)^{-k}\prod_{i=1}^k\lim_{u_i\rightarrow \hat{u}_i}\f{I_i(\hat{u}_1,\dots,\hat{u}_{i-1},u_i)}{(u_i-\hat{u}_i)}\;.
   \end{equation}
If instead $A > k$ we are in the \emph{singular}%
\footnote{This means that more than $k$ singular hyperplanes meet at $\vec{u}=\vec{\hat{u}}$.}
case of JK procedure. The recipe for the singular case in \cite{JK95, Benini:2013xpa} would be problematic for our choice of $\vec{\eta}$. Therefore, we perturb the singularities appearing in eq.~\eqref{eq:to_desing} in the following way:
\begin{multline}\label{eq:desing}
     I_i(u_1,\dots,u_i) \;\mapsto\; \tilde{I}_i(u_1,\dots,u_i) \equiv \\ \equiv
     \f{1}{\jt{u_i - u_{\alpha_{i,1}} + r_{i,1}}}\times\f{\prod_{c_i=1}^{A_i-1}\jt{u_i - u_{\gamma_{i,c_i}} + s_{i,c_i} + \xi_{i,c_i}}}{\prod_{a_i=2}^{A_i}\jt{u_i - u_{\alpha_{i,a_i}} + r_{i,a_i} + \xi_{i,a_i-1}}}\times\\
     \times \prod_{c_i=A_i}^{C_i}\jt{u_i - u_{\gamma_{i,c_i}}+s_{i,c_i}} \times f_i(u_1,\dots,u_i)\;.
   \end{multline}
 We observe that the second factor has neither poles nor zeroes since numerator and denominator vanish simultaneously, by construction. This kind of desingularisation amounts to ``explode'' our pole into $\binom{A}{k} $ non-singular poles. We can number all these poles with a $k$-ple $(\vec{t},\vec{p})\equiv \bigl( (t_1,p_i),\dots,(t_k,p_k) \bigr)$, where $t_i=1,\dots,k$, $p_i=1,\dots,A_i$ and no duplicates $(t_i,p_i)$ are possible. The new poles occur at\footnote{The $(t_i,p_i)$ means that we are using the $p_i$\textsuperscript{th} singular hyperplane of $I_{t_i}$ to determine the intersection point.} $u_i=\hat{u}_i^{(\vec{t},\vec{p})}$ where $\hat{u}_i^{(\vec{t},\vec{p})}$ is such that
   \begin{equation}
     \{\hat{u}_{t_i}^{(t,p)} - \hat{u}_{\alpha_{t_i,p_i}}^{(t,p)} + r_{t_i,p_i}+\xi_{t_i,p_i} = 0\}_{i=1}^k\;,
   \end{equation}
   whose solution, when it exists, is of the form
   \begin{equation}
     \hat{u}_i^{(\vec{t},\vec{p})} = \hat{u}_i + \sum_{i=1}^k\ell_i^{(\vec{t},\vec{p})}\xi_{t_i,p_i}\;,
   \end{equation}
   for certain coefficients $\ell_i^{(\vec{t},\vec{p})}$. Now it is easy to compute the residues
   \begin{equation}
     \res_{\{u_i \rightarrow \hat{u}_i^{(\vec{t},\vec{p})} \}_{i=1}^k}\tilde{I}(u_i,\dots,u_k) = \bigl( 2\pi\eta(\tau)^3 \bigr)^{-k} \lim_{\{u_i \rightarrow \hat{u}_i^{(\vec{t},\vec{p})} \}_{i=1}^k}\prod_{i=1}^k\f{\tilde{I}_i(u_1,\dots,u_i)}{(u_i-\hat{u}_i^{(\vec{t},\vec{p})})}\;,
  \end{equation}
  in the following cases (which are the cases of interest):
  \begin{itemize}
  \item if $(t_i,p_i)=(i,1)$ for\footnote{This is actually the ``unshifted pole'' at $\vec{u}=\vec{\hat{u}}$.} $i=1,\dots,k$ and $A_i=C_i+1$ for all $i=1,\dots,k$ we have:
    \begin{equation}\label{eq:res_des_non_zero}
      \res_{\{u_i \rightarrow \hat{u}_i \}}\tilde{I}(u_i,\dots,u_k) = \bigl( 2\pi\eta(\tau)^3 \bigr)^{-k}\prod_{i=1}^kf_i(\hat{u}_1,\dots,\hat{u}_i)\;;
    \end{equation}
  \item if $(t_i,p_i)\neq (i,1)$ and $A_i = C_i + 1$ for at least one $i=1,\dots,k$ we have
    \begin{equation}
       \res_{\{u_i \rightarrow \hat{u}_i^{(\vec{t},\vec{p})} \}_{i=1}^k}\tilde{I}(u_i,\dots,u_k) = 0\;;
     \end{equation}
   \item if $A_i < C_i + 1$ for at least one $i=1,\dots,k$, for every pole we have
     \begin{equation}
        \res_{\{u_i \rightarrow \hat{u}_i^{(\vec{t},\vec{p})} \}_{i=1}^k}\tilde{I}(u_i,\dots,u_k) = 0\;.
     \end{equation}
  \end{itemize}
  This is because in eq.~\eqref{eq:desing} the numerator and the denominator in the second factor take the same value for $u_i=\hat{u}_i$ by construction, and because if $C_i > A_i-1$ for some $i$ the last factor sets the whole expression to zero. The condition $A_i=C_i+1$ for every $i$ means that the order of singularity of the integrand is $1$ for every $u_i$. If this condition is satisfied, we saw that, after this desingularisation procedure, only the ``unshifted pole'' (i.e. $\vec{u}=\vec{\hat{u}}$) gives non-zero contribution and this contribution is independent of the desingularisation parameters $\xi$'s. This means that once the pole is selected by JK condition, no matter if it lies in the regular or singular case, after the (possibly required) desingularisation procedure, it yields one and just one contribution. Moreover, eq.~\eqref{eq:res_des_non_zero} suggests also a very simple way to evaluate residues provided we have $A_i= C_i + 1$ for all $i=1,\dots,k$: it implies that we have to evaluate $[2\pi\eta^3(\tau)]^{-k}I(\vec{u})$ at $\vec{u}=\vec{\hat{u}}$ simply dropping from it all factors (in the numerator as well as in the denominator) that vanish at this point, as we did in eq.~\eqref{eq:zmid}. In this way the result is both finite and non-zero.

  As a final comment we observe that of all these $\binom{A}{k}$ regular poles, into which the singular pole has been exploded, only $\prod_{i=1}^kA_i$ respect the JK condition. They are the ones corresponding to $t_i$s all different among each other. As far as the opposite case is concerned, in fact a matrix of charges containing two columns like
  \begin{equation}
    \begin{pmatrix}
   & &  & * & * & & & \\
   & &  & \vdots & \vdots & & & \\
   & &  & * & * & & &\\
   \phantom{*}&   \cdots &\phantom{*} & 1 & 1 &\phantom{*} & \cdots &\phantom{*} \\
   & &  & 0 & 0 & & &\\
   & &  & \vdots & \vdots & & & \\
   & &  & 0 & 0 & & &
    \end{pmatrix}
  \end{equation}
  cannot be put in the form \eqref{eq:Qfinal} by swapping rows and columns since in \eqref{eq:Qfinal} there are no couples of $1$'s in the same raw. This last observation will be  useful in the following subsection.

\subsection{Plane partition construction}
\label{app:pp}

In this section we prove that the only set of $U_{(l,m,n)}$ as in eq.~\eqref{eq:pole_location} yielding a non-vanishing JK residue are those in correspondence with plane partitions. In particular, these contributions come from the poles satisfying $\mathscr{S}_k=k$, where $\mathscr{S}_k=C-A$ at rank $k$. This is consistent with the results obtained in the previous subsection. Notice that in this case we can compute residues thanks to eq.~\eqref{eq:res_des_non_zero}. We proceed in the proof by induction on $k$. The case $k=1$ is trivial: the only pole we have is at $u=0$ and the only box representing it is $U_{(1,1,1)}$; clearly, it is a plane partition and, according to the definition, it is the only plane partition we can form with just one box; in addition we have $C=1$ and $A=0$. Then we suppose that we have already built a plane partition of order\footnote{We write $|\{(l,m,n)\}|=k$ to indicate that the cardinality of the set of indices $(l,m,n)$ we are considering is $k$.} $k$, $\ms{U}_k\equiv\{U_{(l,m,n)}\}_{|\{(l,m,n)\}|=k}$ and see what happens when we ``add a box'', $U_{(l',m',n')}$ so that we have the new arrangement $\ms{U'}_{k+1} = \ms{U}_k \cup U_{(l',m',n')}$. ``Adding a box'' means, at the level of integral \eqref{elliptic genus N=1}, that we are spotting the poles of the integrand of $Z_{k+1}^{(1)}$ once we have already classified the poles of the integrand of $Z_{k}^{(1)}$. Our claim is that $\mathscr{S}_{k+1} = \mathscr{S}_k +1$ if $\ms{U}'_{k+1}$ is again a plane partition while, if the new arrangement is not a plane partition, its residue is trivially zero. Once this claim is proved we have the correspondence stated above by induction on $k$.

Let us prove the claim. We distinguish two main cases to organise the proof. Consider the case in which $U_{(l',m',n')}\not\in\ms{U}_k$, which in terms of boxes means that $U_{(l',m',n')}$, the new box, does not coincide with another box in $\ms{U}_k$. In order to increase the singularity, we see from Table~\ref{tab: sing} there are four possibilities: either $(l',m',n')=(a+1,b,c)$ or $(l',m',n')=(a,b+1,c)$, or $(l',m',n')=(a,b,c+1)$ or $(l',m',n')=(a-1,b-1,c-1)$, where $U_{(a,b,c)}\in \ms{U}_k$. We treat the first three possibilities together as a first case and the last possibility as a second case. 

Let us now introduce some useful terminology and notation: for practical reason it is convenient to denote $l'_1\equiv l'$, $l'_2\equiv m'$ and $l'_3\equiv n'
$, moreover we define\footnote{Explicitly $\vec{e}_1=(1,0,0)$, $\vec{e}_2=(0,1,0)$ and $\vec{e}_3=(0,0,1)$.} $\vec{e}_i$ ($i=1,2,3$) directions, as the direction along which the plane partition increases, corresponding to $\epsilon_i$. We will call the ``direction (and orientation) of a face'' of the boxes, the direction (and orientation) of the unit vector normal to this face, pointing outward the box. Thus, every box in the plane partition has three external faces (EFs), which are the ones whose orientation is aligned\footnotemark with one of the $\vec{e}_i$, and three internal faces (IFs), which are the ones whose orientation is anti-aligned\footnotemark[\value{footnote}] with one of the $\vec{e}_i$. \footnotetext{For aligned we mean same direction and same orientation while for antialigned we mean same direction but different orientation.} We will say that a face is free if it is not in common with any other boxes (there is no boxes attached there). 

Let's start the proof in the first case.
The box $U_{(l',m',n')}$ can have either $0$, $1$, $2$ o $3$ free IFs:
\begin{itemize}
\item If there are $3$ free IFs this mean that the box sits in the origin and we have already considered that case $k=1$;
\item If there are $2$ free IFs, let us suppose\footnote{The other cases are easily obtained by permuting $1$, $2$ and $3$.} that they have direction $-\vec{e}_1$ and $-\vec{e_2}$ while the face which is not free have direction $-\vec{e}_3$. Since, by inductive hypothesis, we have the box $U_{(l_1,l_2,l_3-1)}$ in the plane partition, there is one poles arising from a singular hyperplane of type\footnote{We recall that the name of singular and zero hyperplane are listed in Table~\ref{tab: sing}.} $H_A^{(3)}$. Then we can make the following distinction:
  \begin{itemize}
  \item if $l'_1=l'_2=1$ the new arrangement is by definition a plane partition. There are neither source of zeroes nor other sources of poles. So $\Delta\mathscr{S}\equiv \mathscr{S}_{k+1}-\mathscr{S}_k=1$;
  \item if $l'_1=1$ but $l'_2\neq 1$ we do not have a plane partition. In this case there is a zero from $Z_A^{(23)}$ since the box $U_{(l',m'-1,n'-1)}$ is present. There are no other source of poles. We have therefore $\Delta\mathscr{S}\leq 0$;
    \item if $l'_1\neq 1$ and $l'_2\neq 1$ the new arrangement is not a plane partition. In this cases the following boxes are present: $U_{(l',m'-1,n'-1)}$, $U_{(l'-1,m',n'-1)}$ from which we get two zeroes ($Z_A^{(23)}$ and $Z_A^{(13)}$) and $U_{(l'-1,m'-1,n'-1)}$ from which we get a pole thanks to $H_V$. There are not any other source of poles. So we have $\Delta\mathscr{S}\leq 0$.
    \end{itemize}
  \item If there is $1$ free IF, let us suppose that it has direction $-\vec{e}_1$ and that the direction of non-free IF are $-\vec{e}_2$ and $-\vec{e}_3$. Then we have the following boxes: $U_{(l',m'-1,n')}$ and $U_{(l',m',n'-1)}$, which give us two poles (from $H_A^{(2)}$ and $H_A^{(3)}$) and $U_{(l',m'-1,n'-1)}$ which gives a zero (from $Z_A^{(23)}$). Then we can distinguish the following subcases:
    \begin{itemize}
    \item if $l'_1=1$ the new arrangement is a plane partition. There are neither sources of poles nor sources of zeroes; then $\Delta\mathscr{S}=1$;
      \item if $l'_1\neq 1$ we have several boxes to consider: from $U_{(l'-1,m'-1,n'-1)}$ we have a pole (from $H_V$), while from $U_{(l'-1,m'-1,n-)}$, $U_{(l'-1,m',n'-1)}$ and $U_{(l',m'-1,n'-1)}$ we have zeroes (from $H_A^{(12)}$, $H_A^{(13)}$ and $H_A^{(23)}$). There are no more source of poles. Then $\Delta\mathscr{S}\leq 0$.
      \end{itemize}
      \item If there are not free IFs, this means that we have several boxes: $U_{(l'-1,m',n')}$, $U_{(l',m'-1,n')}$, $U_{(l',m',n'-1)}$ from which we get three poles (from $H_A^{(1)}$, $H_A^{(2)}$ and $H_A^{(3)}$), another pole from $U_{(l'-1,m'-1,n'-1)}$ (from $H_V$), while from $U_{(l'-1,m'-1,n)}$, $U_{(l'-1,m',n'-1)}$ and $U_{(l',m'-1,n'-1)}$ we have zeroes (from $H_A^{(12)}$, $H_A^{(13)}$ and $H_A^{(23)}$). Then $\Delta\mathscr{S}=1$.
\end{itemize}

We have now to consider the second case in which $(l',m',n')=(a-1,b-1,c-1)$ for some $U_{(a,b,c)}\in \ms{U}_k$. Since we want $U_{(l',m',n')}\not\in\ms{U}_k$, at least one among $a$ or $b$ or $c$ must be equal to $1$. The hyperplane $H_V$ provide us a pole, then:
\begin{itemize}
\item if $l'_1=l'_2=l'_3=1$, there is a zero from $Z_F$, so $\Delta\mathscr{S}=0$;
  \item if, suppose, $l_1\neq 1$ then we have the box $U_{(l'-1,m',n')}$ that gives a zero by $Z_A^{(23)}$. So $\Delta\mathscr{S}=0$.
\end{itemize}

This exhausts the way one can add $U_{(l',m',n')}\not \in \ms{U}_k$ to $\ms{U}_k$.  Until now we proved that if $\ms{U}_{k+1}$ is a plane partition $\Delta\mathscr{S}=1$ and so the residue computed in this case is not zero. We have finally to examine what happens if we add a box $U'_{(l',m',n')}$ which coincides with another box $U_{(l',m',n')}$ of $\ms{U}_k$.

Using the notation of the previous subsection,%
\footnote{It is always possible to order boxes $U_{(l',m',n')}$ first by $l'$ then by $m'$ and lastly by $n'$. In this way one finds the corresponding $\hat{u}_i$. With this choice the set $\{\hat{u}_i\}_{i}^s$ is a plane partition for $s=1,\dots,k$.}
if one takes some $\hat{u}_{i'} = \hat{u}_i$, the ordering \eqref{eq:to_desing} will be of the form
  \begin{multline}
    I(\vec{u}) = I_1(u_1){\cdot}\dots{\cdot} I_{i}(u_1,\dots,u_i)\times \\ \times I_{i'}(u_1,\dots,u_i,u_{i'})I_{i+1}(u_1,\dots,u_i,u_{i'},u_{i+1}){\cdot}\dots{\cdot} I_k(u_1,\dots,u_k)\;.
  \end{multline}

  Now we can desingularise $I(\vec{u})$ and get $\tilde{I}(\vec{u})$. Now let us examine the following product
  \begin{equation}
    \tilde{I}_1(u_1){\cdot}\dots{\cdot}\tilde{I}_i(u_1,\dots,u_i) \, \tilde{I}_{i'}(u_1,\dots,u_i,u_{i'})\;,
  \end{equation}
  we will have that $A_j = C_j +1$ for $j=1,\dots,i$ and also for $j=i'$. Then, from the integrand \eqref{elliptic genus N=1} we have that $\tilde{I}_{i'}$ contains a term which is $\theta_1^2(\tau| u_i-u_{i'})$, and therefore vanishes when one take the residue w.r.t. the ``unshifted pole'' $u_i=u_{i'}=\hat{u}_i=\hat{u}_{i'}$. From this we conclude that an arrangement of boxes in which two of them occupy the same place do not give contribution.

This proves that the number of the fundamentals charge vector in $\ms{Q}$ can just be $f=1$ and so there is only one block.

\bibliography{EnumerativeGeometry}
\end{document}